\documentclass[11pt,a4paper]{emulateapj}

\usepackage{psfrag}
\usepackage{psfig}
\usepackage{pspicture}
\usepackage{graphicx}

\usepackage{amssymb,amsmath}
\usepackage{natbib}

\def\apj{ApJ}

\journalinfo{\@Accepted for publication in the \apj}
\submitted{Accepted for publication in the \apj}

\begin{document}

%\title {Physical Properties of Star-forming Galaxies in Filaments \& the Field at \lowercase{$z\sim$}0.5: Spectroscopic Evidence for Reduced Electron Density and Enhanced Metallicity in Filament Galaxies}
%\title {Spectroscopic Study of Star-forming Galaxies in Filaments \& the Field: Evidence for Reduced Electron Density and Enhanced Metallicity in Filament Galaxies}
\title {Spectroscopic Study of Star-forming Galaxies in Filaments and the Field at \lowercase{$z\sim$}0.5: Evidence for Environmental Dependence of Electron Density}

\author{
Behnam Darvish,\altaffilmark{1}
Bahram Mobasher,\altaffilmark{1}
David Sobral,\altaffilmark{2,3,4}
Shoubaneh Hemmati,\altaffilmark{1}
Hooshang Nayyeri,\altaffilmark{5}
Irene Shivaei \altaffilmark{1}
}

\setcounter{footnote}{0}
\altaffiltext{1}{University of California, Riverside, 900 University Ave, Riverside, CA 92521, USA; email: bdarv001@ucr.edu}
\altaffiltext{2}{Instituto de Astrof\'{\i}sica e Ci\^encias do Espa\c co, Universidade de Lisboa, OAL, Tapada da Ajuda, PT 1349-018 Lisboa, Portugal}
\altaffiltext{3}{Departamento de F\'{\i}sica, Faculdade de Ci\^encias, Universidade de Lisboa, Edif\'{\i}cio C8, Campo Grande, PT1749-016 Lisbon, Portugal}
\altaffiltext{4}{Leiden Observatory, Leiden University, P.O. Box 9513, NL-2300 RA Leiden, The Netherlands}
\altaffiltext{5}{University of California, Irvine, Irvine, CA 92697, USA}

\begin{abstract}

We study the physical properties of a spectroscopic sample of 28 star-forming galaxies in a large filamentary structure in the COSMOS field at $z\sim$0.53, with spectroscopic data taken with the Keck/DEIMOS spectrograph, and compare them with a control sample of 30 field galaxies. We spectroscopically confirm the presence of a large galaxy filament ($\sim$ 8 Mpc), along which five confirmed X-ray groups exist. We show that within the uncertainties, the ionization parameter, equivalent width (EW), EW versus specific star-formation rate (sSFR) relation, EW versus stellar mass relation, line-of-sight velocity dispersion, dynamical mass, and stellar-to-dynamical mass ratio are similar for filament and field star-forming galaxies. However, we show that on average, filament star-forming galaxies are more metal-enriched ($\sim$ 0.1$-$0.15 dex), possibly due to the inflow of the already enriched intrafilamentary gas into filament galaxies. Moreover, we show that electron densities are significantly lower (a factor of $\sim$17) in filament star-forming systems compared to those in the field, possibly because of a longer star-formation timescale for filament star-forming galaxies. Our results highlight the potential pre-processing role of galaxy filaments and intermediate-density environments on the evolution of galaxies, which has been highly underestimated.

\end{abstract}


\section{Introduction} \label{intro1}

Metal enrichment in galaxies is one of the most important aspects of galaxy evolution as it incorporates the fundamental quantities affecting their evolution such as inflows of cold gas, feedback processes and outflows (stellar winds, supernovae and AGN activity), stellar mass, gas fraction, star-formation rate (SFR), and environment. A relatively tight correlation between metallicity and stellar mass (the mass-metallicity relation) has been investigated, with a redshift evolution reflecting the metal enrichment of galaxies over cosmic time \citep{Tremonti04,Erb06,Maiolino08,Xia12,Foster12,Sanchez13,Henry13,Zahid13,Yabe14,Steidel14,Masters14,Sanders15,Maier15}. It has been shown that this mass-metallicity relation originates from a more fundamental mass, metallicity, SFR relation (the fundamental metallicity relation), forming a tight surface in this 3D space \citep{Mannucci10,Lara-lopez13,Stott13,Bothwell13,Maier14,Nakajima14,delosReyes15}. The mass-metallicity evolution is therefore attributed mostly to the selection of galaxies with higher SFRs at higher redshifts \citep{Mannucci10}. Simulations and physical models suggest that the metal content of galaxies is the result of a balance between inflows and outflows set by the mass outflow rate, whereas the gas content is regulated by the gas inflow rate and the gas consumption timescale within the galaxy, governed by the star-formation law \citep{Dave11,Lilly13}. 

Several studies have scrutinized the potential role of environment in the evolution of the mass-metallicity relation, showing that environment has a second-order, insignificant effect in the mass-metallicity relation, with the majority of studies indicating a slight metallicity enhancement in denser environments \citep{Mouhcine07,Cooper08,Ellison09,Hughes13,Sobral13,Kulas13,Shimakawa15,Kacprzak15,Sobral15,
Valentino15}. However, these studies focus mostly on the cluster versus field as two extreme regions in the distribution of galaxies or rely on the local number density of galaxies as a proxy for their environment. The distribution of galaxies is not just limited to clusters and the field as they naturally belong to the complex network of the cosmic web, comprising clusters, groups, filaments, walls, and voids. Using local number density of galaxies as a probe of their environment has its own drawbacks as well. For example, regions with similar local densities such as galaxy groups, the outskirts of clusters, and galaxy filaments might host different physical processes. As an example of complexities that may arise by focusing only on clusters or local densities, \cite{Ellison09} have shown that cluster galaxies are on average more metal enriched than the field, but this enhancement is primarily driven by local overdensity of galaxies and not merely cluster membership. Therefore, a better understanding of the evolution of galaxies can be achieved by considering a broader range of environments, e.g., by including galaxy filaments in studies.
 
Filamentary structures are among the most interesting objects in astronomy. On galactic and circumgalactic scales, galaxies are primarily cold-gas fed through narrow filamentary streams as shown in simulations \citep{Dekel09a}. Recent observations are starting to reveal the existence of such feeding filaments \citep{Cantalupo14,Martin14a,Martin14b,Martin15}, with sizes up to a few hundred kpc, possibly leading to star-formation in star-forming and post-starburst galaxies \citep{Lilly13,Nayyeri14,Schaye15}. The smooth versus clumpy nature of these filamentary streams has a significant effect on the formation and evolution of disk- and bulge-dominated galaxies \citep{Dekel09a,Dekel09b}. Optical emission line filaments with typical sizes of a few tens of kpc are observed in many cooling flow galaxy clusters, often extending from the central brightest cluster galaxy that is accreting the cooling gas \citep{Ford79,Cowie83,Hu85,Heckman89,Crawford99,McDonald09,McDonald10,McDonald11}.       
On larger scales (a few to several Mpc), filaments make the backbone of the cosmic web of galaxies \citep{Bond96}, linking clusters, groups, and denser regions in the density field. Although they are expected to occupy $\lesssim$ 10\% of the volume of the cosmic web, simulations have shown that galaxy filaments contain most of the mass in the universe ($\sim$ 40\%, \citealp{Aragon-Calvo10}) and host a significant fraction of baryons in the form of a warm-hot intergalactic medium gas in the temperature range 10$^{5}$-10$^{7}$ K \citep{Cen99,Dave01,Cen06,Dave10,Klar12}. Filaments are seen in the optical wavelengths as the large-scale thread-like concentration of galaxies (e.g., \citealp{Pimbblet04,Scoville07b,Kovac10,Guzzo13,Scoville13,Alpaslan14,Tempel14,Darvish14,Darvish15}), in X-ray as a warm-hot gas in emission and absorption (e.g., \citealp{Scharf00,Zappacosta02,Kaastra03,Finoguenov03,Nicastro05a,Nicastro05b,Werner08,Danforth10,
Zappacosta10,Nicastro10}), in the infrared as possible sites of enhanced star-formation activity, linking clusters (e.g., \citealp{Koyama08,Biviano11,Coppin12,Pintos-Castro13}), and in the weak-lensing studies as a bridging distribution of dark matter, connecting galaxy clusters (e.g., \citealp{Jauzac12,Dietrich12,Higuchi15}).
 
Despite the significant importance of filaments, their observational aspect is still poorly investigated or often ignored (see e.g., \citealp{Kodama01,Ebeling04} for some earlier studies on filaments). Nevertheless, the limited number of studies in this field has already highlighted the importance of filaments as regions with an enhanced fraction of star forming galaxies and/or star-formation activity \citep{Porter07,Porter08,Koyama08,Fadda08,Biviano11,Coppin12,Pintos-Castro13,Darvish14}. There are also some systematic surveys/works trying to expand galaxy evolution studies beyond the cluster realm by probing the much larger scale structures (of the order of $\sim$ 10 Mpc and beyond) around massive galaxy cluster candidates (see e.g., \citealp{Kodama05,Kartaltepe08,Lubin09,Sobral11}). In an attempt to study the role of filaments in galaxy evolution, \cite{Darvish14} investigated the star-formation activity in a large-scale structure (LSS, $\sim$ 10 $\times$ 15 Mpc) at $z\sim$0.84 in the COSMOS field \citep{Scoville07} by identifying a prominent filamentary structure, traced by the distribution of narrow-band selected H$\alpha$ star-forming galaxies \citep{Sobral11}. They showed that although, on average, the SFR, SFR$-$mass relation, and specific star-formation rate for star-forming galaxies show similar trends in different environments, the fraction of H$\alpha$ emitting galaxies is enhanced in filaments compared to clusters and the field, possibly due to mild galaxy$-$galaxy interactions. However, a thorough grasp of the role of filaments in galaxy evolution can only be obtained through a detailed spectroscopic study of galaxies in different environments including filaments (see e.g., \citealp{Tanaka07,Tanaka09}).

In this paper, we focus on the spectroscopic study of galaxies in a significantly large filament ($\sim$ 8 Mpc) at $z\sim$0.53 in the COSMOS field and compare it with a control sample of field galaxies at the same redshift. We will investigate the physical properties of galaxies such as electron density, metallicity, ionization parameter, equivalent width (EW), velocity dispersion, and dynamical mass in these two environments and scrutinize possible differences.

The format of this paper is as follows. In Section \ref{LSS-c}, we briefly describe the methods used to extract the LSS and the filament. Section \ref{obs} deals with the spectroscopic observations. Section \ref{SF} presents the star-forming emission-line galaxies selected for the spectroscopic analysis. Our results are given in Section \ref{result5} and discussed in Section \ref{dis5}. We provide a summary of the results in Section \ref{Sum}.

Throughout this work, we assume a flat $\Lambda$CDM cosmology with H$_{0}$=70 kms$^{-1}$ Mpc$^{-1}$, $\Omega_{m}$=0.3 and $\Omega_{\Lambda}$=0.7. All magnitudes are expressed in the AB system and stellar masses and star formation rates are given assuming a Chabrier IMF \citep{Chabrier03}.

\begin{figure*} 
 \begin{center}
\includegraphics[width=7.0in]{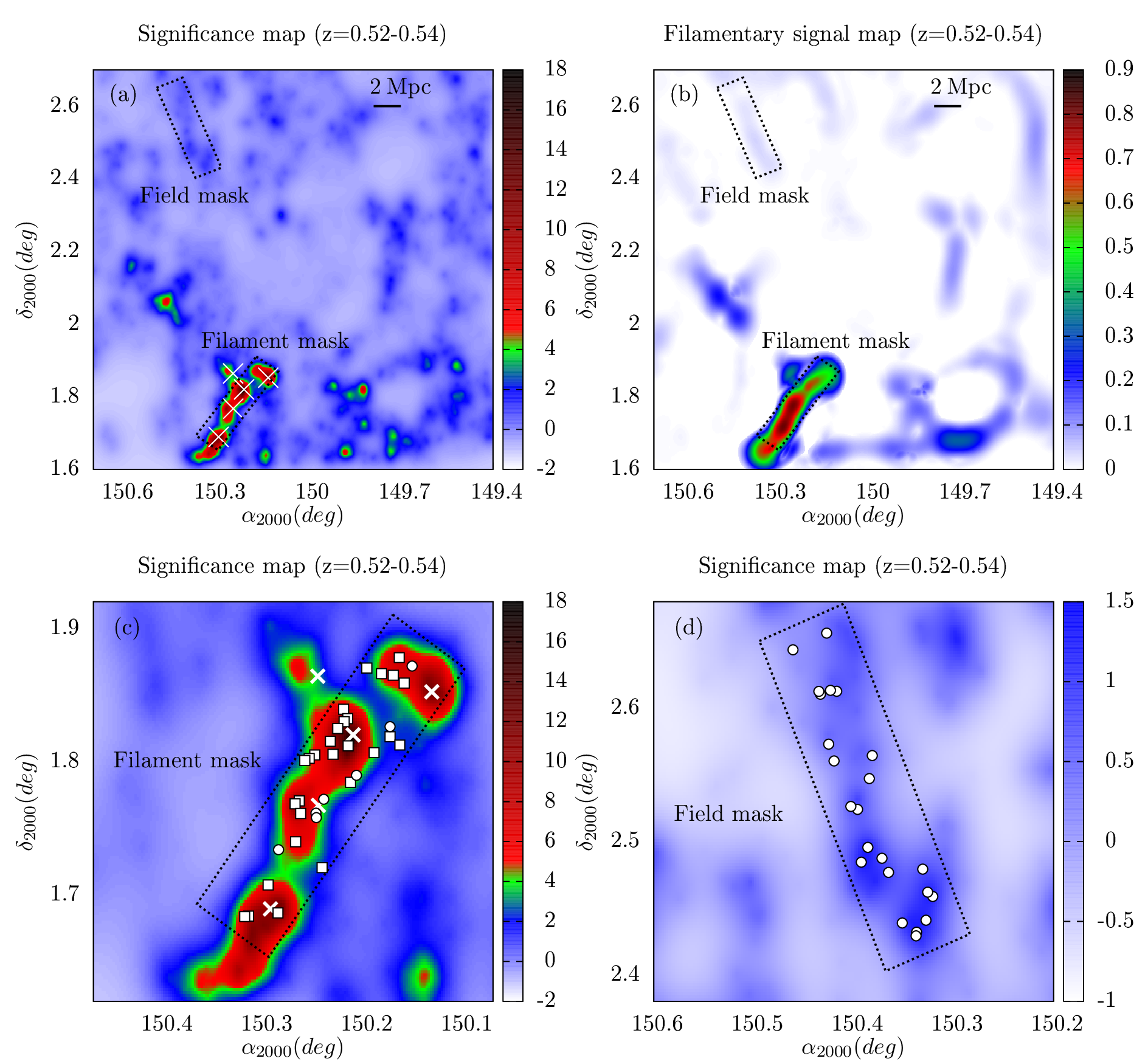}
\caption{(a): Significance map of the density field in the COSMOS at $z$=0.52$-$0.54, estimated via the weighted adaptive kernel density estimator explained in \cite{Darvish15}. A large-scale structure ($\sim$ 8 Mpc) with a filament-like morphology is clearly seen in the lower left. There are at least five confirmed X-ray groups \citep{Finoguenov07,George11} lying along, or in the vicinity of this filamentary structure (white crosses). Black dashed regions show the approximate position of filament and field masks, designed for spectroscopy (Section \ref{strat}). (b): Filamentary signal map, constructed using the MMF algorithm \citep{Aragon-Calvo07,Darvish14} at $z$=0.52$-$0.54, reveals a large filamentary signal values ($>$ 0.5) in the region covering the LSS, further supporting the reality of the filament. (c): A close-up view of the significance map, showing the filamentary LSS, the approximate position of the filament mask, and the projected position of the filament star-forming, emission-line galaxies (white squares) selected for our spectroscopic study (Section \ref{SF}). Note that some galaxies on the filament mask are selected as the field sources (white circles) as their spectroscopic redshift is outside the redshift range selected for the filament membership (Section \ref{SF}). (d): A close-up view of the significance map, showing the approximate position of the field mask, and the projected position of the field star-forming, emission-line galaxies (white circles) selected for our spectroscopic study (Section \ref{SF}).}
\label{fig:LSS1}
 \end{center}
\end{figure*}

\section{Motivation and the LSS Extraction} \label{LSS-c}

We use the density field estimation of \cite{Darvish15} in the COSMOS field \citep{Scoville07} to search for large overdensities of galaxies that have a filamentary morphology. The surface density field was estimated in narrow $z$-slices, using a K$_{s}$ $<$ 24 \citep{McCracken12} sample of galaxies with accurate photometric redshifts (photo-$z$ uncertainty $\Delta z\lesssim$0.01 at $z\sim$0.5, \citealp{Ilbert13}), with the aid of the robust weighted adaptive kernel density estimator. The photo-$z$ probability distribution function (PDF) of individual galaxies were carefully incorporated into the density field estimation in order to minimize the projection effect. At $z\sim$0.53, we find a filament-like, significant overdensity of galaxies. Figure \ref{fig:LSS1}(a), shows the estimated density field in the COSMOS field at $z\sim$0.52-0.54. A large-scale structure with a thread-like morphology is evidently seen in the lower left side of Figure \ref{fig:LSS1}(a), with a significance of $\sigma\sim$8$-$18 \footnote{Significance is defined as $\sigma$=($\Sigma -\Sigma_{bg}$)/$\sigma_{bg}$, where $\Sigma$ is the estimated surface density, $\Sigma_{bg}$ is the mean surface density, and $\sigma_{bg}$ is the standard deviation around the mean surface density values.}. The same structure can also be seen in the density maps produced by \cite{Scoville13} at $z\sim$0.53.

To further investigate the reality of this structure, we used the 2D version of the Multi-Scale Morphology Filter (MMF) algorithm \citep{Aragon-Calvo07,Darvish14}, which is able to disentangle the cosmic web into its components such as filaments and clusters. This algorithm describes the local geometry of each point in the density field based on the signs and the ratio of eigenvalues of the Hessian matrix (second order derivative of the density field). For example, if the local geometry of a point resembles a thread-like structure, this algorithm results in a large filamentary signal for the point. The filamentary signal value (a value between 0 and 1) shows the degree of resemblance to a filament, with higher values indicating more resemblance. We apply the MMF algorithm to our density field at $z\sim$0.53, which results in large filamentary signal values ($>$ 0.5) in the region covering the LSS (Figure \ref{fig:LSS1}(b)).

The reality of this filamentary structure is further supported by the X-ray group catalogs of \cite{Finoguenov07} and \cite{George11} in the COSMOS field. There are at least five confirmed intermediate-density X-ray groups lying along, or in the proximity of this filamentary structure (white crosses in Figures \ref{fig:LSS1}(a) and \ref{fig:LSS1}(c)), in agreement with the expectations from the cosmic web simulations that galaxy groups/clusters are connected through the network of filaments \citep{Bond96}. The X-ray groups are in the redshift range $\delta z$=0.527$-$0.530, with virial masses M$_{200}$=1.8$-$4.6$\times$10$^{13}$ M$\odot$ \citep{George11}. This filamentary structure has a physical length of $\sim$ 8 Mpc and it might be even longer as it is located near the edge of the COSMOS field. According to simulations \citep{Aragon-Calvo10}, such long filaments have $<$ 10\% chance of being found in the density field, making them unique targets for the studies of the cosmic web and the LSS.

\section{Spectroscopic Observations} \label{obs} 
 
\subsection{Target Selection for Spectroscopy} \label{target}

In selecting the sources for spectroscopy, we define two samples. The first sample comprises the potential filament members and the second is a control sample of field galaxies at the same redshift. The main filament targets are selected to be in the vicinity of the filament, brighter than i$^{+}$ $<$ 22.5 and to have $>$ 50\% of their photo-$z$ PDF lying within the redshift slice $\Delta z\sim$0.52$-$0.54 centered at $z$=0.53. The width of the redshift slice is $\sim$2 times the median photo-$z$ uncertainty at $z\sim$0.5.
The main field targets are selected similarly to those in the filament, except that they are located in a random point in the COSMOS field with no sign of overdensities. We also consider fillers as we design multi-object masks for spectroscopy. The filler sources on the filament mask are selected to be in the vicinity of the filament, brighter than i$^{+}$ $<$ 22.5, with their photo-$z$ in the range $\Delta z$=0.41$-$0.52 or $\Delta z$=0.54$-$0.65. The filler sources on the field mask are chosen in a similar way as that of the filament fillers except that they are located in the random position of the field mask. In selecting the sources (main targets and fillers, in filament and the field), we discard those flagged as potential AGN, based on the publicly available catalog of AGN candidates in the COSMOS field \citep{Trump07}. Therefore, we expect little to no AGN contamination.

\begin{table*}
\begin{center}
\caption{Observing Log and Spectroscopic Mask Properties} 
\begin{scriptsize}
\centering
\begin{tabular}{lccccccc}
Mask Name & Date & R.A. & Dec. & Position Angle & Exp. Time & Number of Sources\\
& & J2000 & J2000 & deg & ks \\
\hline
Filament & 28 Dec. 2014 & 10:00:59.36 & $+$1:47:04.4 & $-$42.0 & 5.4 & 52 targets, 46 fillers \\
Field & 29 Dec. 2014 & 10:01:35.78 & $+$2:32:08.0 & $+$27.0 & 2.8 & 17 targets, 50 fillers \\
\hline
\label{log}
\end{tabular}
\end{scriptsize}
\end{center}
\end{table*}

\begin{figure*}
\centering
\includegraphics[width=7in]{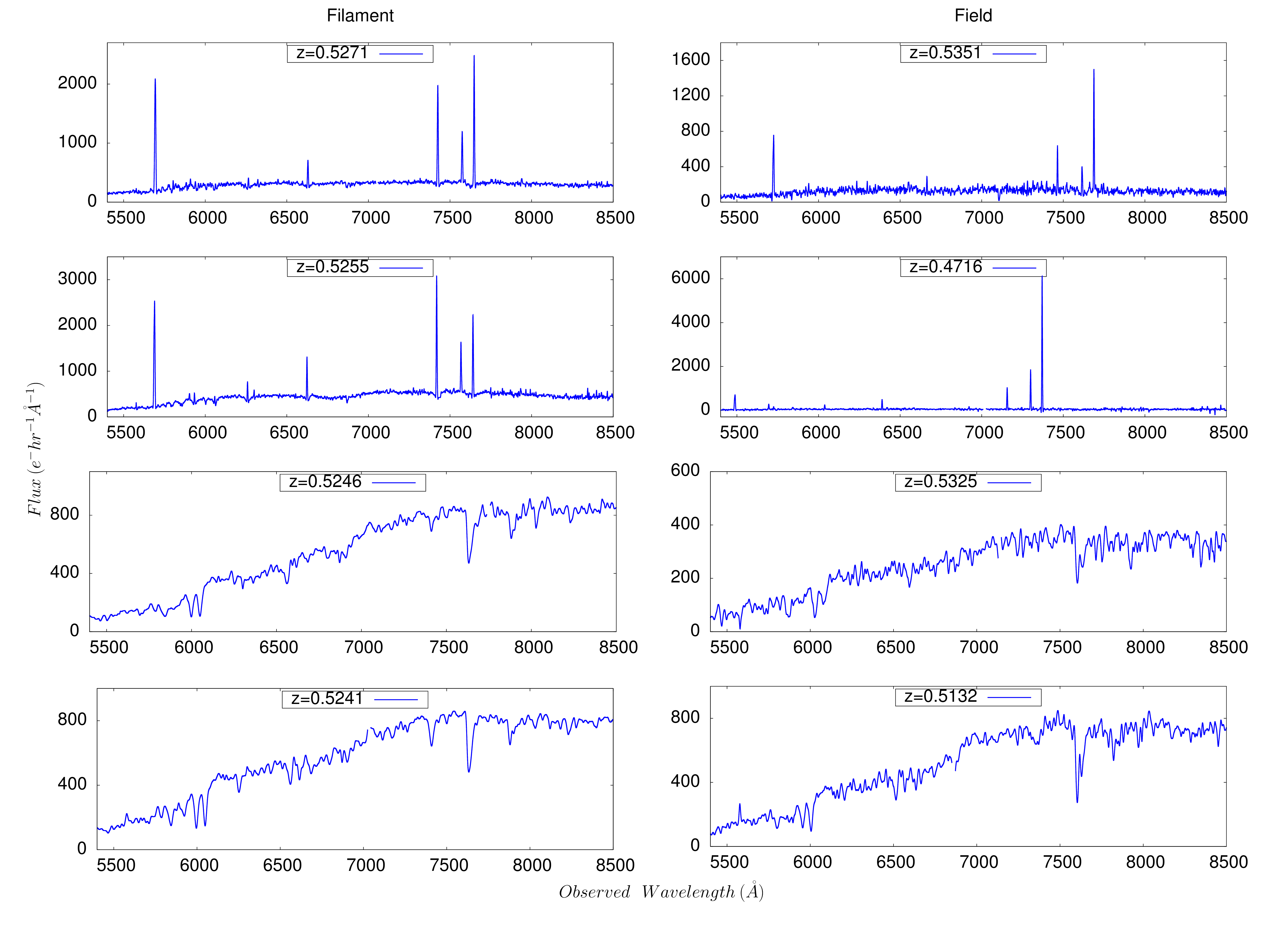}
\caption{Some examples of the observed spectra, including quiescent and star-forming galaxies, in the filament and the field masks in the wavelength range 5400$-$8500 {\AA}. The labels show the extracted redshift of the samples. Note the quality (high S/N) of the spectra.}
\label{fig:spec}
\end{figure*}   

\subsection{Observational Strategy} \label{strat}

Spectroscopic observations were made on 28 and 29 December 2014 with the DEIMOS spectrograph on the Keck II telescope under clear conditions. We used a 600 line/mm diffraction grating (0.65 {\AA}/px dipersion) with a spectral span of $\sim$5300 {\AA} and a GG455 order blocking filter. Slit width is set to 0.75" for all sources. We select the central wavelength of $\lambda_{c}$=7200 {\AA} which gives us an observed wavelength coverage $\sim$4550$-$9850 {\AA}. Since the majority of the targets are expected to be at $z\sim$0.53, we achieve the rest-frame wavelength coverage $\sim$3000$-$6400 {\AA}, given that all the slits are located at the center of the mask. This enables us to cover major spectroscopic features that can be potentially used to estimate SFR (e.g. [OII]$\lambda$3727, H$\beta$), metallicity ([OII]$\lambda$3727, [OIII]$\lambda$4959, [OIII]$\lambda$5007, H$\beta$), electron density ([OII]$\lambda\lambda$3726,3729 doublet), and ionization parameter ([OII]$\lambda$3727, [OIII]$\lambda\lambda$4959,5007) for the main targets. We designed two masks, one containing the filament targets and fillers and the other containing the control sample of field galaxies (targets and fillers). The masks were designed in such a manner to maximize the number of main targets (Black dashed regions in Figure \ref{fig:LSS1}).

The filament mask consisted of 52 main targets and 46 fillers and the field mask contained 17 main sources and 50 fillers \footnote{There is also one Lyman-$\alpha$ candidate in the filament and one in the field mask. The spectroscopic confirmation and analysis of these Lyman-$\alpha$ targets are presented in \cite{Matthee15} and \cite{Sobral15b}.}. The filament mask was observed on December 28 for 5.4 ks exposure time (4 individual exposures of 1.2 ks and one exposure of 0.6 ks), with the midpoint airmass of $\sim$1.1 and an average seeing of $\sim$0.5". We observed the field mask on December 29 for $\sim$2.8 ks total exposure time, with the midpoint airmass of $\sim$1.05 and an average seeing of $\sim$0.7". We note that after the spectroscopic redshift extraction, some fillers may turn out to be filament members or some fillers might be used as field galaxies. We will use these extra possible sources to increase the sample size for our analysis. Table \ref{log} summarizes the observing log.

\subsection{Data Reduction and Calibration} \label{calib}

In order to reduce the DEIMOS data, we use the $spec2d$ package \citep{Cooper12,Newman13}. The reduction process involves flat fielding, cosmic ray removal, sky subtraction and wavelength calibration on a slit-by-slit basis. We used standard Kr, Xe, Ar, and Ne arc lamps for wavelength calibration. No dithering pattern was used for sky subtraction. The pipeline also extracts the 1D spectrum from the reduced 2D one per slit using the extraction algorithm of \cite{horne86}, which relies on the sum of fluxes at each wavelength in an optimised window around the 2D spectrum. We flux-calibrate the 1D spectra using the standard stars $Feige 34$ and $Hz 44$. The half-light radius of the sources is in the range $\sim$0.2$-$0.9". Therefore, we lose some flux at the 0.75" slit width due to the larger angular size of some of the sources and primarily, due to the seeing. We correct the flux-calibrated spectrum of each source (for the slit-loss) by scaling it with the observed photometry of the source. This is performed by integrating the spectra over the Subaru filter response profiles in broad ($r$ and i$^{+}$), intermediate ($IB574$, $IB624$, $IB679$, $IB709$, $IB738$, $IB767$ and $IB827$), and narrow ($NB711$ and $NB816$) band filters and finding the required slit-loss correction term at each waveband. The median (estimated with the Hodges$-$Lehmann estimator) of the slit-loss correction scaling terms at these bands for each individual source is used as its slit-loss correction term. $Subaru$ multi-waveband magnitudes are extracted from the publicly available optical catalog of the COSMOS field \citep{Capak07}. Figure \ref{fig:spec} shows examples of selected spectra in the filament and the field.

\subsection{Spectroscopic completeness} \label{comp}

The overall spectroscopic completeness for the main targets in the filamentary structure and the field are $\sim$74\% and $\sim$77\%, respectively. The spectroscopic completeness varies slightly with magnitude. For the main targets with i$^{+}\leq$21.5, we obtain a spectroscopic completeness of $\sim$76\% (filament) and $\sim$73\% (field). At i$^{+}>$21.5, the spectroscopic completeness becomes $\sim$72\% (filament) and $\sim$82\% (field). We obtain a spectroscopic success rate of $\sim$94\% (89\% for fillers) and 100\% for the main targets in the filament and the field, respectively. The spectroscopic sampling rate is generally a function of the 2D angular coordinates (over the DEIMOS field of view). However, due to the small sample size, we do not try to quantify it. We note that it is relatively uniform over the filament and the field regions. 

\subsection{Filament Spectroscopic Confirmation} \label{fil}

\begin{figure}
 \begin{center}
    \includegraphics[width=3.5in]{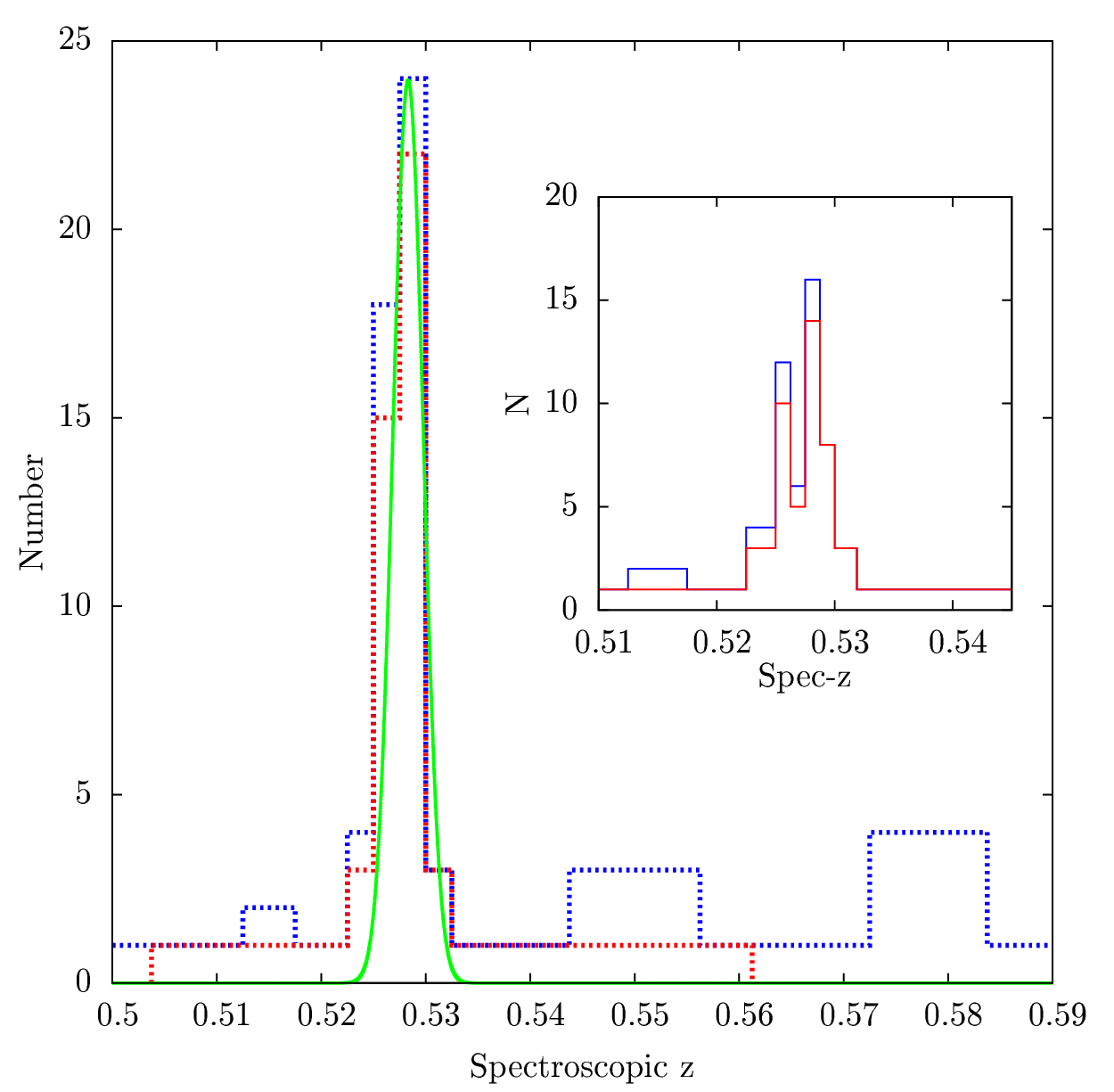}
     \caption{Spectroscopic redshift distribution in the range 0.50$\leq z\leq$0.59 for sources in the filament mask. Dashed blue line shows all the sources (main targets and fillers) in the filament mask, whereas the dashed red line indicates the main filament targets only. A peak at $z\sim$0.53 is seen, confirming the presence of the filament. Solid green line shows a Gaussian fit to the obtained redshift distribution of the main targets, peaking at $z_{peak}$=0.5283, with standard deviation $\sigma$=0.0015. The inner plot shows two individual nearby peaks at $z_{1}$=0.5258$\pm$0.0008 and $z_{2}$=0.5287$\pm$0.0004, showing the existence of possible substructures along the filament.}
\label{fig:z-dist}
 \end{center}
\end{figure}

We visually extract the spectroscopic redshifts using the $SpecPro$ code \citep{Masters11}. The main filament targets yield spectroscopic redshifts in the range $z_{spec}$=0.5067-0.5582 with a very sharp peak at $z\sim$0.53 (see Figure \ref{fig:z-dist}). The filament fillers are spread in the range $z_{spec}$=0.4085$-$0.7797. Figure \ref{fig:z-dist} shows the spectroscopic redshift distribution in the range 0.50$\leq z\leq$0.59 for all the sources (main targets and fillers) in the filament mask (dashed blue line), as well as the main filament targets only (dashed red line). We clearly see a peak at $z\sim$0.53, confirming the presence of the filamentary structure. The average spectroscopic redshift for the filament main targets is $z_{avg}$=0.5279, with standard deviation $\sigma_{z}$=0.0073. We fit a Gaussian function to the obtained redshift distribution of the main targets (solid green line). The Gaussian fit results in a peak redshift of $z_{peak}$=0.5283, with standard deviation $\sigma$=0.0015. 

There is evidence for at least two individual nearby peaks centered at $z_{1}\sim$0.528 and $z_{2}\sim$0.526, as seen in the inner plot of Figure \ref{fig:z-dist}, indicating the existence of substructures along the filament. Fitting a double peaked Gaussian to the distribution of filament main targets yields redshifts $z_{1}$=0.5258$\pm$0.0008 and $z_{2}$=0.5287$\pm$0.0004 for these substructures. The projected 2D distribution of these two nearby peaks also shows two substructures along the filament. Note that there are only 4 ($\sim$8\%) main filament targets that yield spectroscopic redshifts outside the range 0.52$<z<$0.54 originally adopted for main target selection, indicating a successful target selection process and a robust LSS extraction \citep{Darvish15}.
 
\section{Selection of Star-forming, Emission-line Galaxies} \label{SF}

In this work, we compare physical, chemical and dynamical properties of galaxies in the filament and the field that rely on strong emission lines. Therefore, we need to define a sample of star-forming emission line galaxies. We carefully investigate the obtained spectra and select those that visually have clear star-forming features, high quality, high Signal-to-Noise [OII], [OIII], and H$\beta$ emission lines. We also make sure that filament sources have spectroscopic redshifts in the range $z_{spec}$=0.52$-$0.54 and field sources in the range $z_{spec}$=0.47$-$0.59. As we already mentioned in Section \ref{strat}, after the spectroscopic redshift extraction of the sources, some fillers turned out to be filament members and some fillers can be used as field galaxies. We use these extra sources to increase the sample size of star-forming emission line galaxies. The final star-forming sample comprises 28 galaxies in the filament and 31 galaxies in the field \footnote{We later discard one of the field sources as an AGN (Section \ref{BPT}), reducing the field sample size to 30 sources.}. Figures \ref{fig:LSS1}(c) and \ref{fig:LSS1}(d) show the angular distribution of the filament (white squares) and field (white circles) star-forming, emission-line galaxies used in this study.

\subsection{Stellar Mass Estimation} \label{mass}

Stellar masses were derived by \cite{Ilbert13}, using SED template fitting to the UV, optical, and mid-IR photometry. The synthetic templates were generated using BC03 \citep{Bruzual03}, assuming a Chabrier IMF, three different metallicities, an exponentially declining SFH with different $e$-folding time scales ($\tau$=0.1$-$30 Gyr), and the \cite{Calzetti00} extinction law. Contributions from nebular emission lines were included. Stellar masses are based on the median of the stellar mass PDF, marginalized over all other parameters.

The magnitude cut of i$^{+}$ $<$ 22.5 used for sample selection, introduces a varying stellar mass completeness limit at different redshifts. We use the method developed by \cite{Pozzetti10} to estimate the stellar mass completeness (see also \citealp{Ilbert13,Darvish15}). In this method, we assign a limiting mass to each galaxy which corresponds to the stellar mass that the galaxy would have if its apparent magnitude were equal to the magnitude limit used for sample selection (i$^{+}$ $<$ 22.5). At each redshift, the 95\% mass completeness, for instance, is then equal to the stellar mass for which 95\% of galaxies have their limiting mass below it. At $z\sim$0.5, we estimate the mass completeness to be $\sim$70\%, 90\%, and 95\% at log(M/M$_{\odot}$)$\sim$9.3, 9.6, and 9.7 respectively. In the low-mass bin (log(M/M$_{\odot}$)=9$-$9.5) of our star-forming sample, the weighted average of the stellar mass is log(M/M$_{\odot}$)$\sim$9.3. Therefore, in the lower mass range, our sample is $\sim$70\% complete in stellar mass, while at higher mass bins, we achieve a completeness of $>$ 95\%.

\subsection{SFR Estimation} \label{SFR}

We estimate the SFR based on H$\beta$ and [OII] luminosities. We convert dust-corrected H$\beta$ luminosity (L$_{H\beta}$) to star-formation rate using the theoretical value of L$_{H\alpha}$/L$_{H\beta}$=2.86 for case B recombination (temperature T=10,000 K and electron density $N_{e}$=10$^{2}$ cm$^{-3}$, \citealp{Osterbrock89}) and the relation between H$\alpha$ dust-corrected luminosity and SFR given in \cite{Kennicutt98}, modified for the Chabrier IMF:
\begin{equation}
SFR_{H\beta}(M_{\odot}yr^{-1})=1.27\times10^{-41}L_{H\beta}(ergs^{-1})
\end{equation}
The H$\beta$ line is also affected by the Balmer continuum stellar absorption. For brighter galaxies with clear Balmer absorption feature in their spectrum, we correct the H$\beta$ flux by directly fitting a Gaussian function to the the Balmer absorption and adding the missing flux to the H$\beta$ line. For fainter sources where the stellar absorption is not clearly seen, we take the median EW of the Balmer absorption of brighter galaxies and use it to correct for fainter galaxies. The median EW of the H$\beta$ Balmer absoprtion feature is $\sim$0.8 {\AA}, in agreement with \cite{Zahid11}. Nevertheless, the H$\beta$ stellar absorption only results in a 5\%$-$10\% correction to the H$\beta$ flux, which is smaller or of the order of the uncertainty of the measured line flux itself.

\begin{figure}
 \begin{center}
    \includegraphics[width=3.5in]{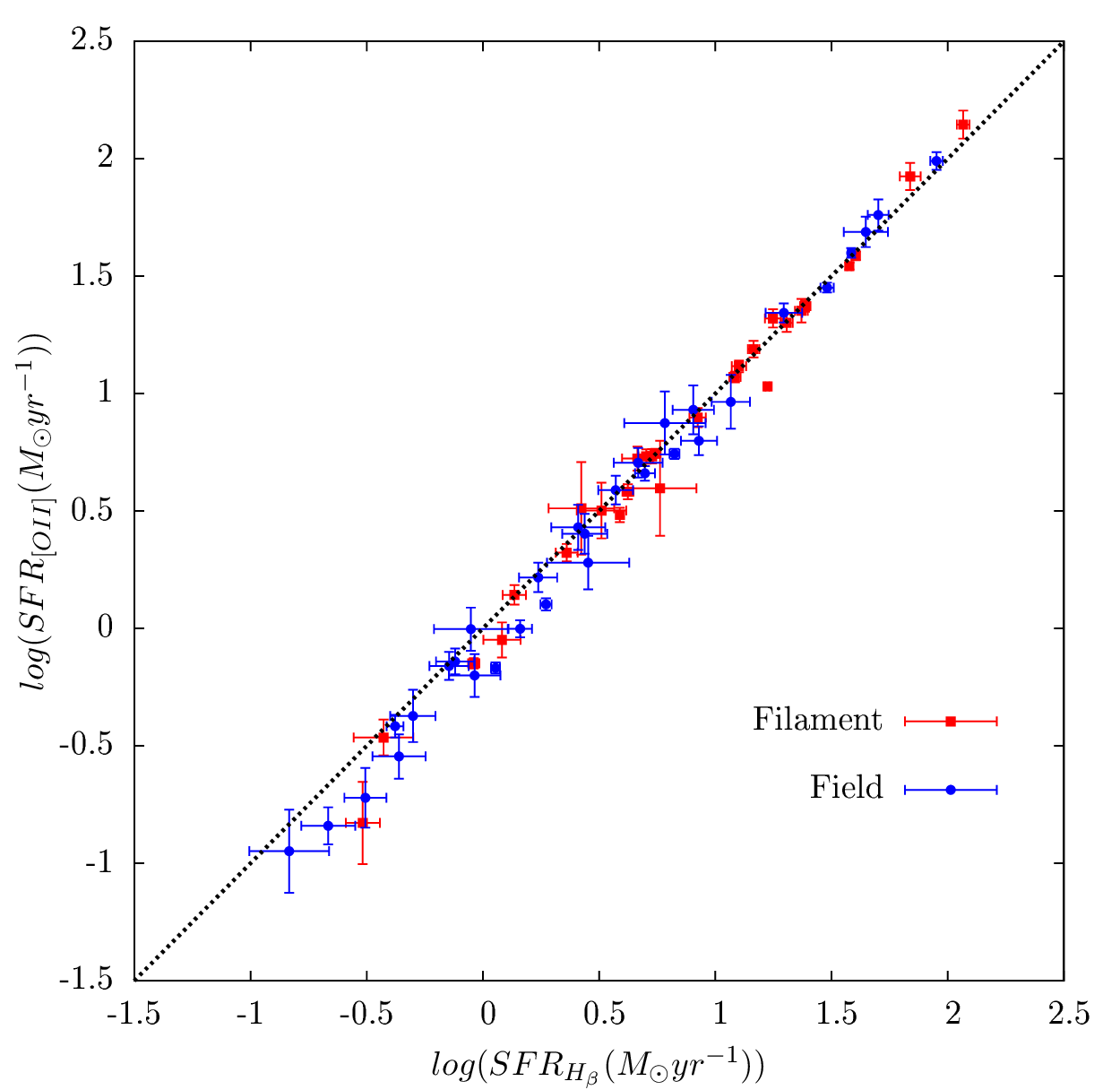}
     \caption{Dust-corrected [OII] versus H$\beta$ SFRs for filament (red) and field (blue) star-forming galaxies. There is a good agreement between the estimated SFRs. The median SFR$_{[OII]}$ and SFR$_{H\beta}$ are consistent to within $\sim$0.01 dex (filament) and $\sim$0.03 dex (field) and the median absolute deviation between the two SFR indicators is only 0.03 dex.}
\label{fig:SFR}
 \end{center}
\end{figure}

The [OII] star-formation rate not only depends on dust-corrected [OII] luminosity but also on metallicity and ionisation state of the gas. Here, we use the metallicity-dependent equation given in \cite{Kewley04}, modified for the Chabrier IMF:\\

\footnotesize
\begin{equation}
SFR_{[OII]}(M_{\odot}yr^{-1})=\frac{4.44\times10^{-42}L_{[OII]}(ergs^{-1})}{(-1.75\pm0.25)[12+log(O/H)]+16.73\pm2.23}
\end{equation}
\normalsize
\\
where 12+log(O/H) is the oxygen abundance metallicity (Section \ref{met}), estimated based on the R$_{23}$ diagnostic, via the R$_{23}$-metallicity relation of \cite{Zaritsky94}:\\
\footnotesize
\begin{equation}
12+log(O/H)=9.265-0.33R_{23}-0.202R_{23}^{2}-0.207R_{23}^{3}-0.333R_{23}^{4}
\end{equation}
\normalsize
\\
where R$_{23}$ metallicity diagnostic is defined as R$_{23}$=([OII]$\lambda$3727+[OIII]$\lambda$4959+[OIII]$\lambda$5007)/H$_{\beta}$.

We dust-correct the emission line fluxes ([OII], [OIII], H$\beta$) using the relation between the reddening of the ionized gas ($E(B-V)_{gas}$) and that of the stellar continuum ($E(B-V)_{star}$) presented in \cite{Calzetti00}:
\begin{equation}
E(B-V)_{gas}=2.27E(B-V)_{star}
\end{equation}
The stellar continuum is estimated in the SED Template-fitting procedure (Section \ref{mass}), assuming the \cite{Calzetti00} attenuation curve. Therefore, the dust-corrected, intrinsic emission line flux ($F_{int}$) is:
\begin{equation}
F_{int}=F_{obs}\times 10^{0.4 k_{\lambda} E(B-V)_{gas}} 
\end{equation}  
where $k_{\lambda}$ is the reddening curve. The relation between stellar and gas extinction is quite uncertain at higher redshifts. While some studies nearly agree with the local universe relation (see e.g., \citealp{Price14} for $z\sim$1.5 galaxies), others have shown that the $E(B-V)_{gas}$=$E(B-V)_{star}$ relation is preferable (see e.g., \citealp{Shivaei15} for $z\sim$2 galaxies and \citealp{Reddy15} for $z\sim$1.4$-$2.6 systems). However, choosing this new relation does not significantly affect many galaxy properties such as metallicity, ionization parameter, electron density, and gas velocity dispersion presented in Section \ref{result5}. Moreover, at the redshift of our study ($z\sim$0.5), \cite{Ly12} have shown that assuming $E(B-V)_{gas}\sim 2E(B-V)_{star}$ for star-forming galaxies results in a better agreement between SED-based and H$\alpha$ SFRs.

Figure \ref{fig:SFR} shows the comparison between dust-corrected [OII] and H$\beta$ SFRs for filament and field samples, exhibiting a very good agreement. The median SFR$_{[OII]}$ and SFR$_{H\beta}$ are consistent to within $\sim$0.01 dex (filament) and $\sim$0.03 dex (field) and the median absolute deviation between the two SFR indicators is only 0.03 dex (filament and field). 
 
We also compare the [OII] and SED-based SFRs (not shown here). This shows a larger dispersion of the order $\sim$0.3 dex. However, this is consistent with e.g., \cite{Steidel14} ($\sim$0.2 dex, H$\alpha$ versus SED SFRs at $z\sim$2$-$3) and \cite{Ciardullo13} ($\sim$0.3 dex, [OII] versus UV SFRs at $z\lesssim$0.5). The source of discrepancy might be due to the stochasticity of the star-formation activity and the difference between the UV and [OII] star-formation timescales ($<$ 20 Myr for [OII] and $\sim$ 100 Myr for UV, \citealp{Kennicutt98}), as the UV SFR is averaged out over a longer timescale. 
  
\begin{figure*}
\begin{center}
\includegraphics[width=7in]{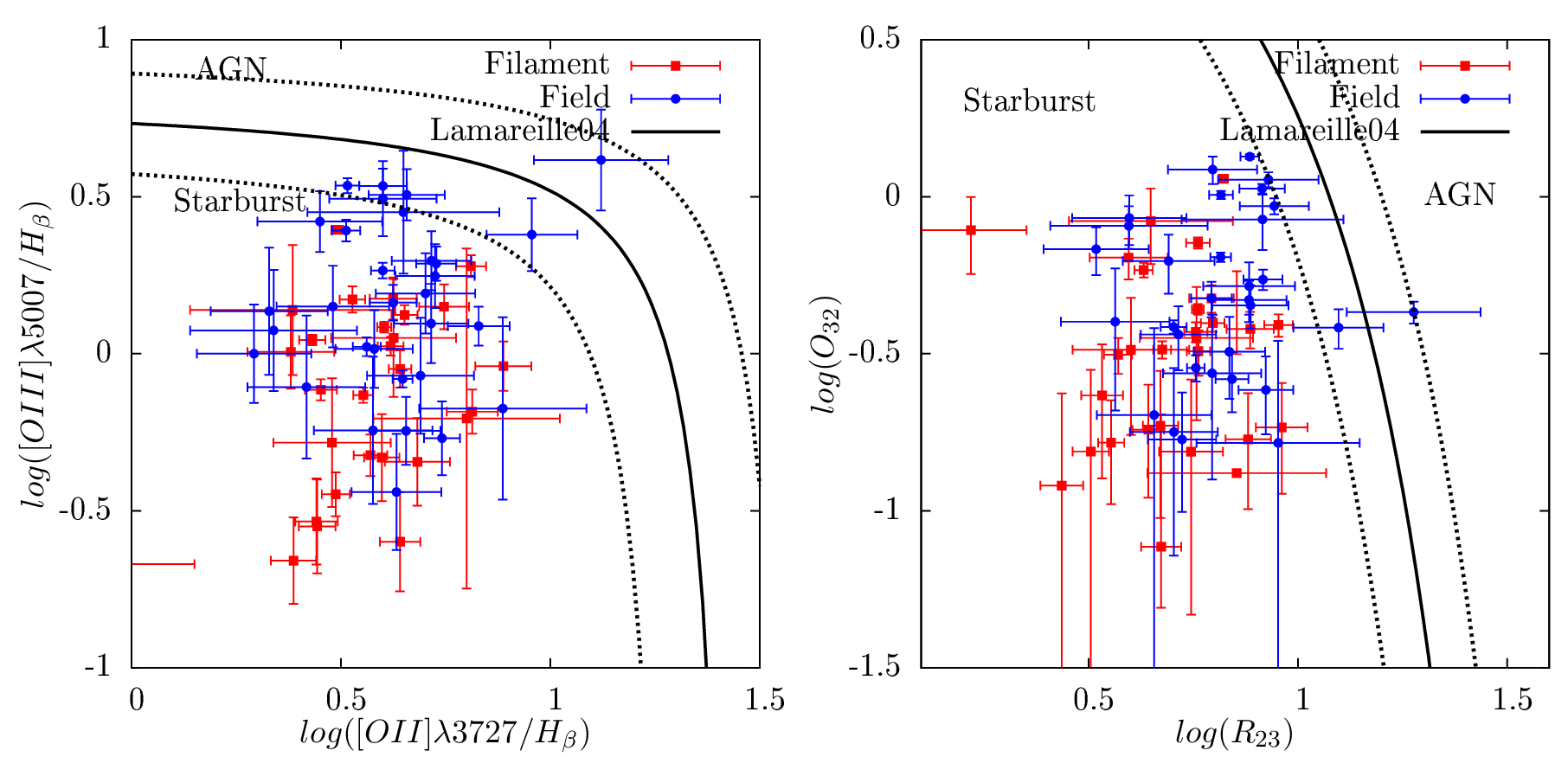}
\caption{[OIII]$\lambda$5007/H$\beta$ versus [OII]$\lambda$3727/H$\beta$ (left) and O$_{32}$ versus R$_{23}$ (right) plots used to separate possible AGN from the filament (red) and field (blue) star-forming galaxies. The separating curve (solid line) is extracted from \cite{Lamareille04} and the dashed lines are $\pm$0.15 dex ($\pm$0.1 dex) uncertainties in the [OIII]$\lambda$5007/H$\beta$ versus [OII]$\lambda$3727/H$\beta$ (O$_{32}$-R$_{23}$) separation curve. Only one source in the field is within the AGN region. We discard this source from our analysis.}
\label{fig:BPT}
\end{center}
\end{figure*}

\subsection{AGN Contamination} \label{BPT}

As we discussed in Section \ref{target}, we originally discarded the potential AGN from the samples selected for spectroscopy \citep{Trump07}. However, we check if the spectroscopic sample is contaminated by any remaining AGN using the available line ratios. Here, we use the [OIII]$\lambda$5007/H$\beta$ versus [OII]$\lambda$3727/H$\beta$ plot, as well as O$_{32}$ versus R$_{23}$ (O$_{32}$ and R$_{23}$ are defined in Section \ref{result5}) plot to separate AGN from star-forming galaxies \citep{Lamareille04}. Figure \ref{fig:BPT} shows the line ratios and the line separating star-forming galaxies from AGN. The separating curve is extracted from \cite{Lamareille04} and the dashed lines are $\pm$0.15 dex ($\pm$0.1 dex) uncertainties in the [OIII]$\lambda$5007/H$\beta$ versus [OII]$\lambda$3727/H$\beta$ (O$_{32}$-R$_{23}$) separation curve. According to Figure \ref{fig:BPT}, all of our sources lie within the star-forming locus except one source. We jettison this source from our analysis in the following sections. We note that there are a few other sources in the field that lie close to the uncertainty curves. These sources do not change the results of this study.

\begin{figure*}
\begin{center}
\includegraphics[height=7in,width=7in]{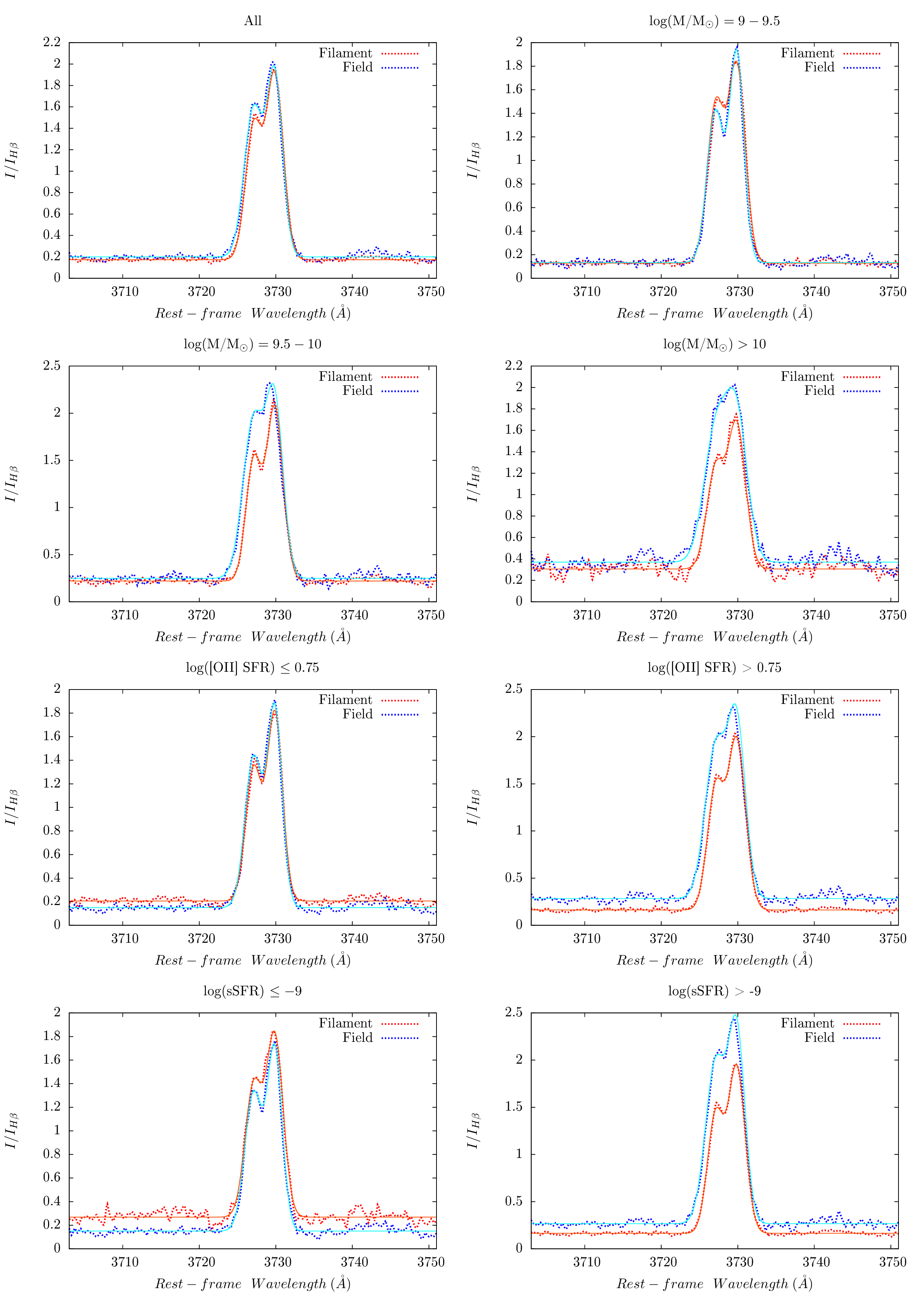}
\caption{Stacked [OII] intensity normalized to that of H$\beta$ at $\lambda_{H\beta}$=4862.7 {\AA} in filament (red dotted line) and the field (blue dotted line) and in stellar mass, SFR, and sSFR bins. We fit the [OII] doublet with double Gaussian functions shown as solid lines. The [OII]$\lambda$3726/[OII]$\lambda$3729 ratios are converted to electron densities, assuming T$_{e}$=10,000 K. Depending on the bin selected, electron densities vary in the range $\sim$ $<$10$-$75 cm$^{-3}$ for filament galaxies and in the range $\sim$ 200$-$740 cm$^{-3}$ for galaxies in the field, significantly lower for filament galaxies.}
\label{fig:OII}
\end{center}
\end{figure*}
  
\subsection{Stacking Procedure}

In order to obtain an average value of the physical quantities presented in Section \ref{result5}, we stack the individual spectra in different environments, as well as stellar mass, SFR, sSFR, and $\mu_{0.32}$ bins (defined in Section \ref{met}). We first shift each flux calibrated spectrum to the rest-frame and normalize it to the flux of the H$\beta$ line at the vacuum wavelength of $\lambda_{H\beta}$=4862.7 {\AA}. We then take the weighted mean of each individual, normalized, rest-framed, flux-calibrated spectrum ($F_{i}$($\lambda$)) by summing them up using:
\begin{equation}
F_{stack}(\lambda)=\frac{\sum\limits_{i=1}^N w_i(\lambda) F_i(\lambda)}{\sum\limits_{i=1}^{N} w_i(\lambda)}
\end{equation}
where $w_{i}(\lambda)$ is the inverse variance of the normalized, rest-framed flux. We define three mass bins (log(M/M$_{\odot}$)=9$-$9.5, log(M/M$_{\odot}$)=9.5$-$10, and log(M/M$_{\odot}$)$>$ 10), two SFR bins (log([OII] SFR)$\leq$0.75 and (log([OII] SFR)$>$ 0.75), two sSFR bins (log(sSFR)$\leq -$9 and log(sSFR)$>$ $-$9), and two $\mu_{0.32}$ bins ($\mu_{0.32}\leq$9.5 and $\mu_{0.32}>$9.5) and will use them in the following section for the analysis of the results.
  
\section{Results} \label{result5}

\subsection{Electron Density} \label{ED}

The average electron density ($N_{e}$) in a nebula can be estimated using the ratio of two lines of the same ion, emitted by different energy levels with almost the same excitation energy. Here, we use [OII]$\lambda$3726/$\lambda$3729 line ratio to estimate the average electron density \citep{Osterbrock89}. Since the [OII]$\lambda$3726 and [OII]$\lambda$3729 lines are not fully separated in our stacked spectra, we fit a double Gaussian function of the following form to the [OII] doublet line flux:
\begin{equation}
f_{\lambda}=(\frac{a_1}{\sqrt{2\pi} \sigma_1}e^{-\frac{(\lambda-\lambda_1)^2}{2\sigma_1^2}}+\frac{a_2}{\sqrt{2\pi} \sigma_2}e^{-\frac{(\lambda-\lambda_2)^2}{2\sigma_2^2}})+f_0
\end{equation}
We fix the parameters $\lambda_1$ and $\lambda_2$ to the vacuum wavelength of [OII] doublet lines, i.e., $\lambda_1$=3727.09 {\AA} and $\lambda_2$=3729.88 {\AA}. The [OII]$\lambda$3726/$\lambda$3729 flux ratio is therefore determined by the ratio of the normalization parameters, $a_1$/$a_2$. We estimate errors of $a_1$ and $a_2$ using a Monte-Carlo simulation. We assume that the true flux at each wavelength follows a Gaussian distribution with a sigma equal to the flux uncertainty. We fit the function $f_{\lambda}$ to the flux values that are randomly selected from their Gaussian distribution at each wavelength. We perform the fitting procedure 100 times and the error of each free parameter is the standard deviation of the 100 estimated values. The [OII]$\lambda$3726/$\lambda$3729 flux ratio is converted to electron density using the $TEMDEN$ task of the $IRAF$ package \citep{DeRobertis87,Shaw95}. We assume an electron temperature of T$_e$=10,000 K as a typical temperature of star-forming regions with near-normal abundances \citep{Osterbrock89}. Figure \ref{fig:OII} shows the stacked [OII] intensity normalized to that of H$\beta$ at $\lambda_{H\beta}$=4862.7 {\AA} in filament and the field and in stellar mass, SFR, and sSFR bins (dotted lines). For comparison, the double Gaussian fits are also shown (solid lines).

We find that the average electron density is significantly lower for star-forming galaxies in the filament compared to that of the field galaxies. The [OII] line ratio is [OII]$\lambda$3726/$\lambda$3729= 0.6909$\pm$0.0044 for the stacked spectra of filament star-forming galaxies, whereas for field star-forming systems, this ratio is [OII]$\lambda$3726/$\lambda$3729=1.0050$\pm$0.0104. These ratios correspond to the average electron densities of $N_e$=22$\pm$4 and 367$\pm$13 cm$^{-3}$ for filament and field galaxies, respectively. The average electron density of the field star-forming galaxies is $\gtrsim$17 times that of star-forming systems in the filament. The average electron density for our filament star-forming galaxies at $z\sim$0.5 is similar to that of the local universe and low-$z$ samples (see e.g., \citealp{Yabe15,Sanders15b}), whereas the typical electron densities in our field sample at $z\sim$0.5 resemble those at higher redshifts (see e.g., \citealp{Steidel14,Masters14,Shimakawa15,Sanders15b,Yabe15}).

\begin{figure}
\begin{center}
\includegraphics[height=7in,width=3.5in]{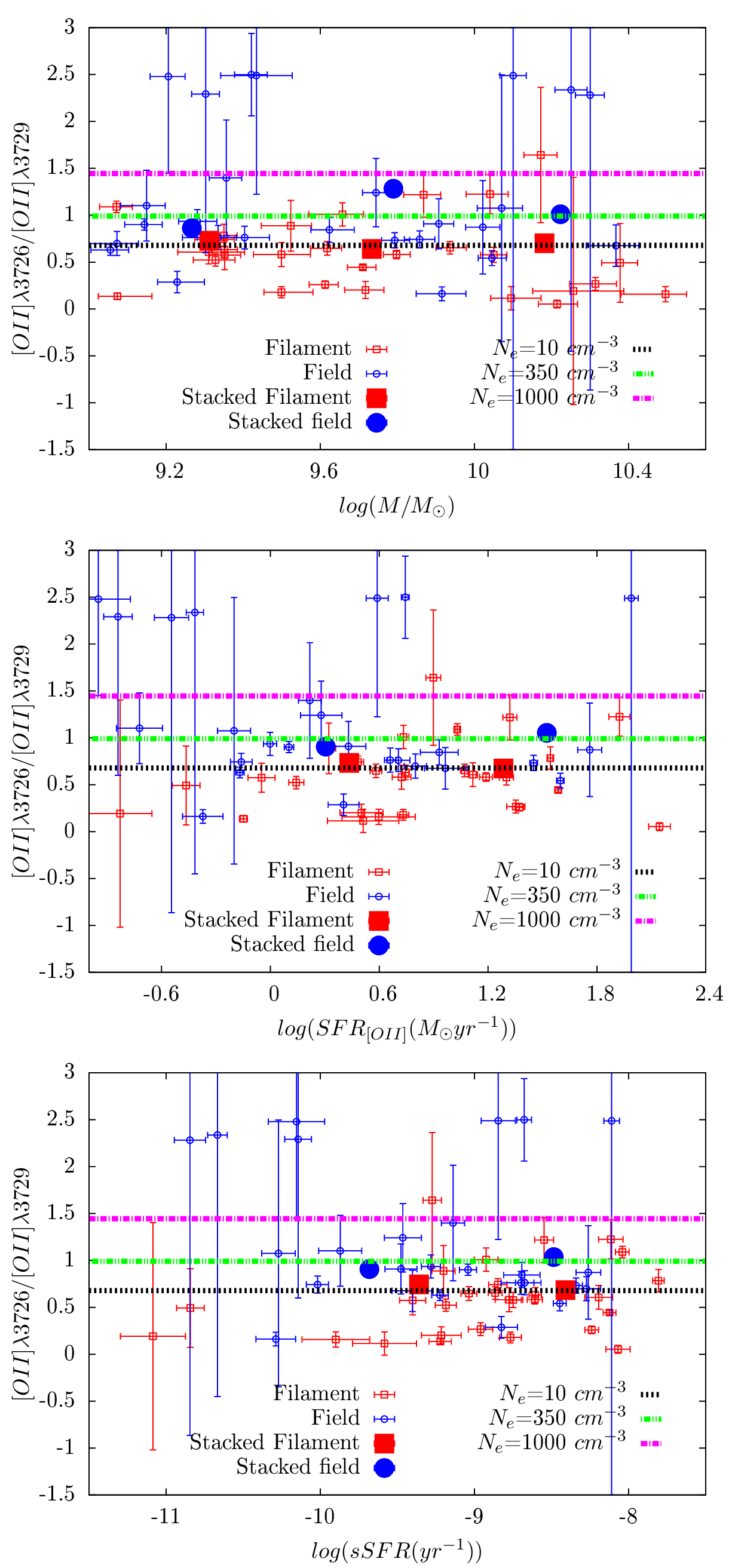}
\caption{[OII]$\lambda$3726/$\lambda$3729 line ratio (as a measure of electron density) for individual filament (red empty points) and field (blue empty points) star-forming galaxies as a function of stellar mass, SFR, and sSFR, along with mass-, SFR-, and sSFR-stacked spectra in filament (red solid points) and the field (blue solid points). The line ratios that correspond to electron densities of $N_{e}$=10, 350, and 1000 cm$^{-3}$ are shown with black, green, and pink dashed lines, respectively. We find no significant relation between the [OII]$\lambda$3726/$\lambda$3729 line ratio (electron density) and stellar mass, SFR, and sSFR for both filament and field star-forming galaxies, with the coefficient of determination $<$0.1 in all cases. However, at a fixed stellar mass, SFR, and sSFR, the average electron density of the stacked spectra for the filament galaxies is significantly lower than that of the field galaxies (compare red solid points with blue ones). Depending on the bin selected, electron densities vary in the range $\sim$ $<$10$-$75 cm$^{-3}$ for filament galaxies and in the range $\sim$ 200$-$740 cm$^{-3}$ for galaxies in the field. The stacked points errorbars are smaller than the size of the points.}
\label{fig:ratio-mass-sfr-ssfr}
\end{center}
\end{figure}

We divide the sample into stellar mass, SFR, and sSFR bins and stack the spectra in each bin. Figure \ref{fig:ratio-mass-sfr-ssfr} shows the [OII]$\lambda$3726/$\lambda$3729 line ratio for individual filament (red empty points) and field (blue empty points) star-forming galaxies as a function of stellar mass, SFR, and sSFR, along with mass-, SFR-, and sSFR-stacked spectra in these two environments. For comparison, we also label the line ratios that correspond to electron densities of $N_{e}$=10 (black dashed line), 350 (green dashed line), and 1000 (pink dashed line) cm$^{-3}$. We find no significant relation between the electron density and stellar mass, SFR, and sSFR for both filament and field star-forming galaxies. We calculate the coefficient of determination (square of the Pearson coefficients) between the [OII]$\lambda$3726/$\lambda$3729 line ratio and stellar mass, SFR, and sSFR for filament and field star-forming galaxies. In all cases, the coefficient of determination is $<$0.1, indicating no clear relation between electron density and stellar mass, SFR, and sSFR. This is fully consistent with e.g., \cite{Yabe15} (for star-forming galaxies at $z\sim$1.4) and \cite{Sanders15b} (for star-forming galaxies at $z\sim$2.3) who found no clear relation between electron density and other properties of galaxies such as stellar mass, SFR, and sSFR.

However, we find that at each given mass, SFR, and sSFR bin, the average electron density of the stacked spectra for the filament galaxies is significantly lower than that of the field galaxies (Figure \ref{fig:ratio-mass-sfr-ssfr}, red and blue solid points). For example, in the mass range of log(M/M$_\odot$)=9$-$9.5, the average electron density is $N_e$=58$\pm$5 (filament) and 201$\pm$10 (field) cm$^{-3}$. For log(sSFR)$>-$9, the average electron density is $N_e$=15$\pm$4 (filament) and 405$\pm$15 (field) cm$^{-3}$, a factor of $\sim$27 larger for field galaxies. In other words, our result regarding the lower average electron density of star-forming galaxies in filament compared to the field does not change if the comparison is performed at fixed stellar mass, SFR, and sSFR bins in these two environments. Depending on the bin selected, electron densities vary in the range $\sim$ $<$10$-$75 cm$^{-3}$ for filament galaxies and in the range $\sim$ 200$-$740 cm$^{-3}$ for galaxies in the field. However, we note that the dispersion in the estimated line ratios (electron densities) for individual galaxies is still large and a larger sample is required to more robustly constrain the environmental dependence of electron densities for star-forming galaxies. The [OII] line ratios, as well as the estimated electron densities in different environments and for different mass, SFR, and sSFR bins of galaxies are given in Table \ref{stack}.

\begin{figure}
\begin{center}
\includegraphics[height=4.67in,width=3.5in]{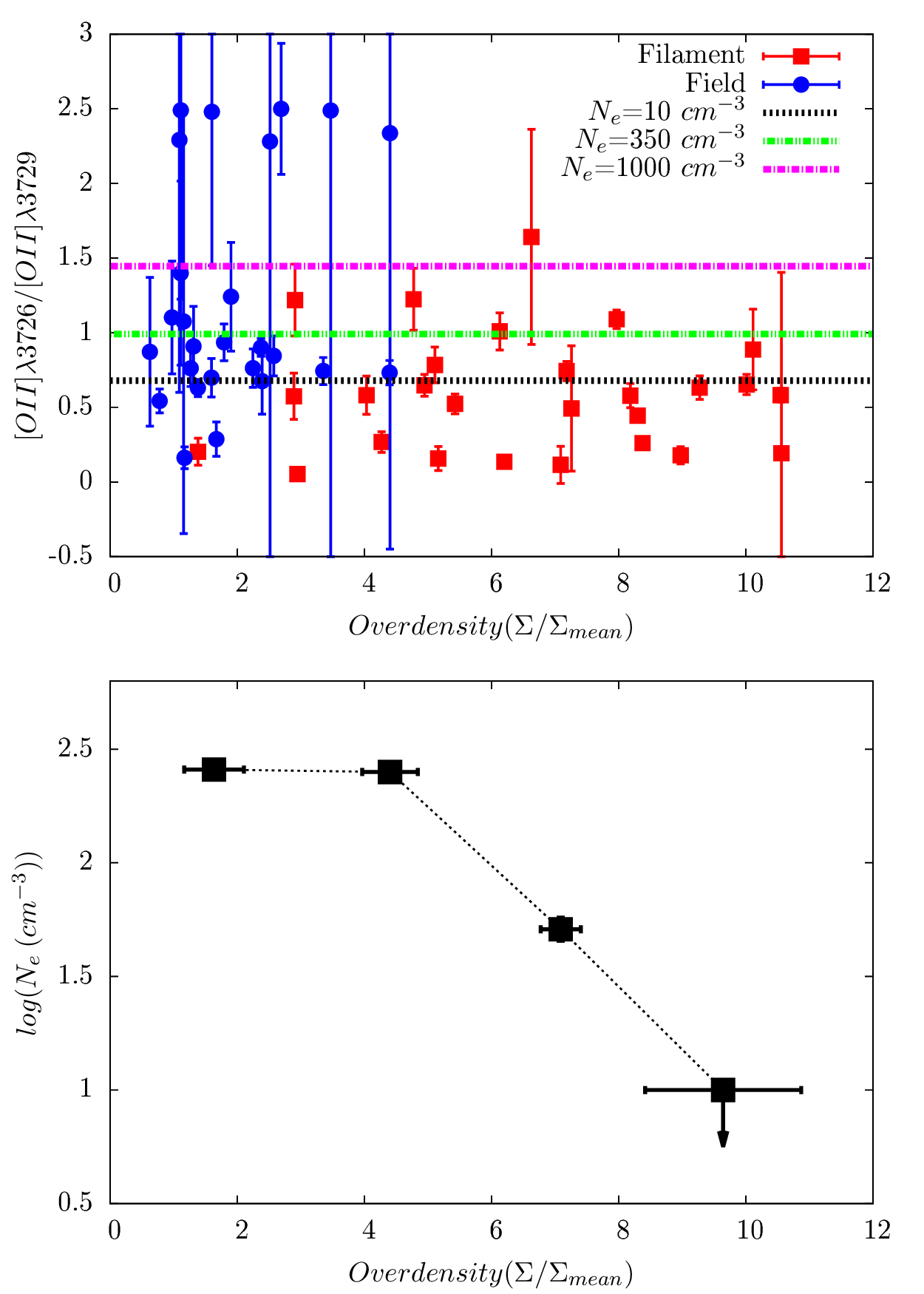}
\caption{Top: [OII]$\lambda$3726/$\lambda$3729 line ratio for filament (red points) and field (blue points) star-forming galaxies as a function of local overdensity of galaxies. The line ratios that correspond to electron densities of $N_{e}$=10, 350, and 1000 cm$^{-3}$ are shown with black, green, and pink dashed lines, respectively. We find that filament star-forming galaxies tend to have lower electron density values. The K-S test result shows that it is significantly likely that the distribution of [OII]$\lambda$3726/$\lambda$3729 line ratios (electron densities) for filament and field star-forming systems are drawn from different parent distributions (p-value=0.000072). Bottom: Average electron densities for the overdensity-stacked spectra as a function of local overdensity of galaxies. We see an anti-correlation between the average electron density and local overdensity of galaxies, suggesting that the environmental dependence of the electron density for star-forming galaxies is possibly a local phenomenon.}
\label{fig:ratio-dens}
\end{center}
\end{figure}

In order to describe the environment of galaxies more quantitatively, we use the local number density of galaxies (estimated in Section \ref{LSS-c}) as a measure of their host environment and investigate the relation between local overdensity of galaxies (overdensity is defined as the estimated local number density divided by the mean local density) and the [OII]$\lambda$3726/$\lambda$3729 line ratios and their corresponding electron densities. Figure \ref{fig:ratio-dens} (top) shows the [OII]$\lambda$3726/$\lambda$3729 line ratio for individual filament (red points) and field (blue points) star-forming galaxies as a function of overdensity values. We also show the line ratios that correspond to electron densities of $N_{e}$=10 (black dashed line), 350 (green dashed line), and 1000 (pink dashed line) cm$^{-3}$. We clearly see that filament star-forming galaxies tend to occupy denser regions and more importantly, tend to have lower electron density values. For example, $\sim$70\% of filament star-forming galaxies have electron densities of $<$10 cm$^{-3}$, whereas only $\sim$15\% field star-forming galaxies have low electron densities of $<$10 cm$^{-3}$. Using the K-S test, we compare the distribution of [OII]$\lambda$3726/$\lambda$3729 line ratios for filament and field star-forming galaxies. The estimated p-value is p=0.000072, indicating that it is significantly likely that the distribution of [OII]$\lambda$3726/$\lambda$3729 line ratios (electron densities) for filament and field star-forming systems are drawn from different parent distributions. We also discard the data points with large line ratio errorbars and reperform the K-S test. The resulting p-value is p=0.0038, showing that the difference between electron densities of filament and field star-forming galaxies is still significant. Furthermore, we stack the galaxies based on their local overdensity value (regardless of whether they are in the filament or the field) and investigate the relation between the electron densities and local overdensity. Figure \ref{fig:ratio-dens} (bottom) shows the estimated electron densities for the overdensity-stacked spectra as a function of overdensity. We clearly see an anti-correlation between the average electron density and the local overdensity of galaxies, especially for $\Sigma/\Sigma_{mean}\gtrsim$6. This suggests that the environmental dependence of the electron density for star-forming galaxies is possibly a local effect.

Using [SII]$\lambda$6716/[SII]$\lambda$6761 line ratio, \cite{Sobral15} found that for a massive post-merger cluster at $z\sim$0.2 \citep{Dawson15,Stroe15}, electron densities are extremely low ($<$ 5 cm$^{-3}$) for cluster star-forming galaxies, much smaller than star-forming galaxies in the outskirts and outside of the cluster. They also found that this is primarily driven by star-forming galaxies in the hottest regions of the cluster, thus likely caused by environmental effects. This is consistent with our result in this section (low to extremely low electron densities for filament galaxies) as we showed that the environment affects the electron densities in a sense that regions with higher local densities host galaxies with lower electron densities. However, the environmental dependence of the electron densities shown in our work seems inconsistent with the recent work of \cite{Kewley15}, who did not find a significant difference between electron densities of cluster and field star-forming galaxies at $z\sim$2. The inconsistency, however, might be due to larger errorbars, smaller sample size, and a different redshift regime compared to our work.
  
\subsection{Metallicity} \label{met}

\begin{figure}
\begin{center}
\includegraphics[width=3.5in]{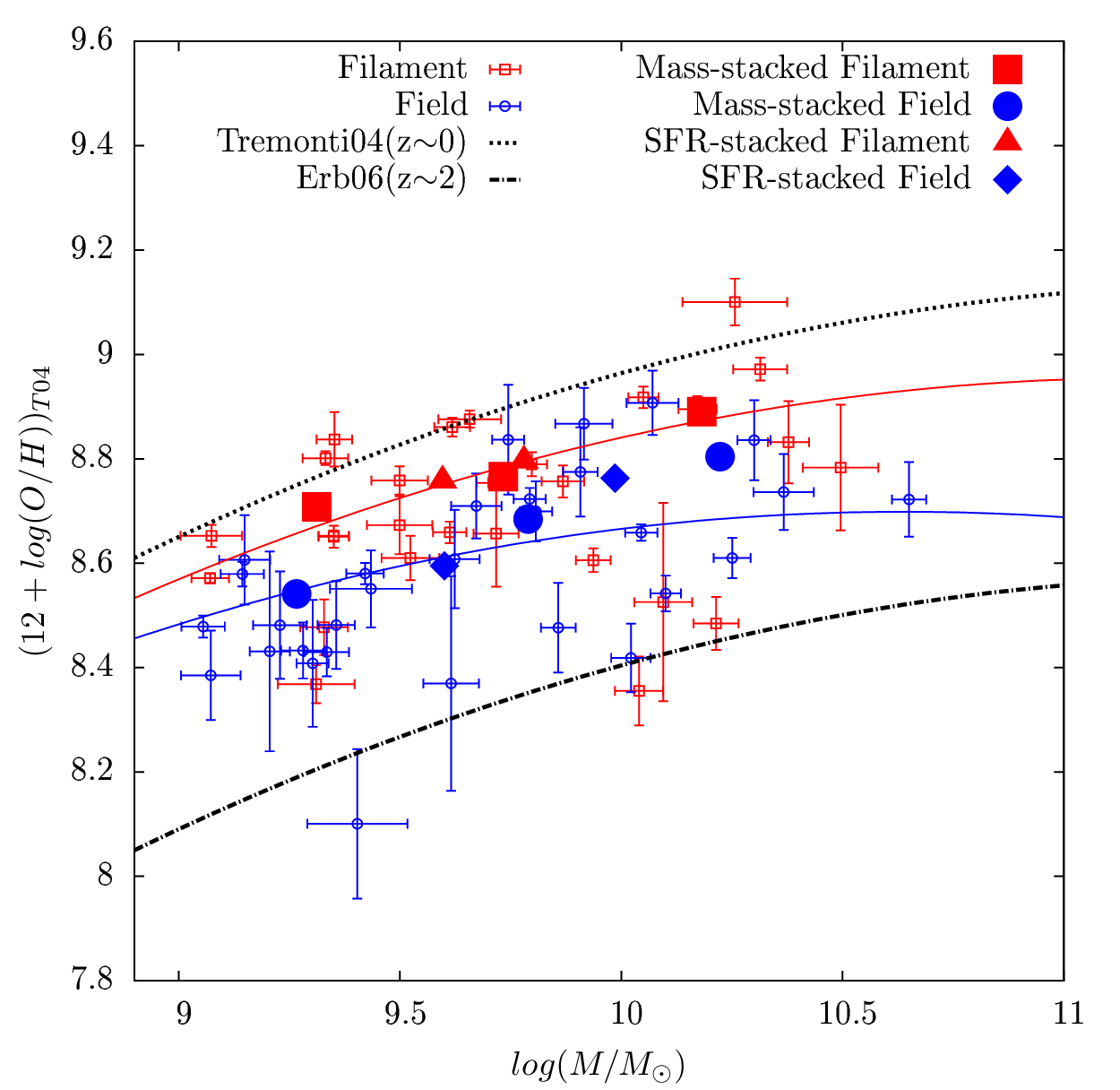}
\caption{Mass-metallicity relation for filament (red empty points) and field (blue empty points) galaxies, along with mass-stacked and SFR-stacked spectra in the filament (red solid points) and the field (blue solid points). Red and blue solid lines represent the fit to the individual galaxies in the filament and the field, respectively. Mass-metallicity relation in the local universe from \cite{Tremonti04} (dotted line) and that of \cite{Erb06} at $z\sim$2 (dashed dotted line) are shown for comparison. The relation lies $\sim$0.06$-$0.13 dex below the local universe trend for filament galaxies and $\sim$0.21$-$0.23 dex for galaxies in the field. Filament galaxies are $\sim$ 0.1$-$0.15 dex more metal-enriched compared to galaxies in the field at a fixed stellar mass bin. The stacked data errorbars are smaller than the size of the points.}
\label{fig:MZ}
\end{center}
\end{figure}

In order to estimate metallicities for filament and field star-forming galaxies, we use the R$_{23}$ index defined as R$_{23}$=([OII]$\lambda$3727+[OIII]$\lambda$4959+[OIII]$\lambda$5007)/H$_{\beta}$ \citep{McGaugh91,Tremonti04,Kewley08}. Here, we use the upper branch of the double-valued R$_{23}$-metallicity relation given by \cite{Tremonti04}:
\begin{equation} \label{eq:EQ-met}
12+log(O/H)=9.185-0.313x-0.264x^{2}-0.321x^{3}
\end{equation}
where $x$=log(R$_{23}$).
\\
Using equation \ref{eq:EQ-met}, we evaluate the average metallicity for the stacked spectra of star-forming galaxies in the filament and the field. The metallicity of the stacked spectra in the filament is 12+log(O/H)=8.766$\pm$0.004, about 0.13 dex larger than that of the field galaxies (12+log(O/H)=8.639$\pm$ 0.007). It is known that the metallicity is also a function of stellar mass and star-formation rate \citep{Tremonti04,Mannucci10} and the metallicity difference between average filament and field star-forming galaxies might be due to different mass and/or SFR of galaxies in these two environments. In order to check this, we divide the star-forming galaxies into mass and SFR bins and stack their spectra. We find that at a given stellar mass bin, star-forming galaxies in the filament are more metal-enriched (by $\sim$0.1$-$0.15 dex) compared to those in the field (Figure \ref{fig:MZ}). For example, at log(M/M$_{\odot}$)=9$-$9.5 mass bin, metallicities are 12+log(O/H)=8.708$\pm$0.004 and 8.541$\pm$0.007 in the filament and the field, respectively. Similar results are obtained at fixed SFR bins. That is, at fixed SFR bins, star-forming galaxies in the filament are more metal-enriched. For example, for log(SFR)$\leq$0.75, star-forming galaxies in the filament are $\sim$0.16 dex more enriched in metals than their field counterparts. Table \ref{stack} gives the metallicities of the stacked spectra for different subsamples.

Figure \ref{fig:MZ} shows the mass-metallicity relation for individual galaxies (red and blue empty points), as well as their stacked results in filament and the field (red and blue filled points). For comparison, we also show the local universe mass-metallicity relation from \cite{Tremonti04} (modified for stellar masses based on the Chabrier IMF) and that of the \cite{Erb06} at $z\sim$2 ($-$0.56 dex below the $z\sim$0 trend). Our obtained mass-metallicity relation at $z\sim0.5$ lies between $z\sim$0 and $z\sim$2 trends. For filament galaxies, it lies $\sim$0.06$-$0.13 dex below the local universe trend, whereas for galaxies in the field, the difference is $\sim$0.21$-$0.23 dex. We fit the metallicity of individual data points in the filament and the field with a function of the form:
\begin{equation}
12+log(O/H)=c(t-m_{0})^{2}+z_{0}
\end{equation}
where $t$=log(M/M$_{\odot}$). We fix the parameter $c$ to that of \cite{Tremonti04} relation ($c$=$-$0.08026). The fitting is performed using the robust nonlinear least-squares Marquardt-Levenberg algorithm. The estimated parameters are $m_{0}$=11.192$\pm$0.339 and $z_{0}$=8.955$\pm$0.094 in the filament and $m_{0}$=10.639$\pm$0.236 and $z_{0}$=8.698$\pm$0.041 in the field. The intrinsic metallicity scatter (after subtracting in quadrature the median observational uncertainties in measured metallicities) around the mean mass-metallicity relation is $\sim$0.17 dex (filament) and $\sim$0.13 dex (field). The results of the fits are shown in Figure \ref{fig:MZ}. The fits to individual data points also show difference ($\sim$ 0.1$-$0.2 dex) between metallicities in filament and the field at a given stellar mass, although the dispersion around the fitted lines is large. 

\begin{figure}
\begin{center}
\includegraphics[width=3.5in]{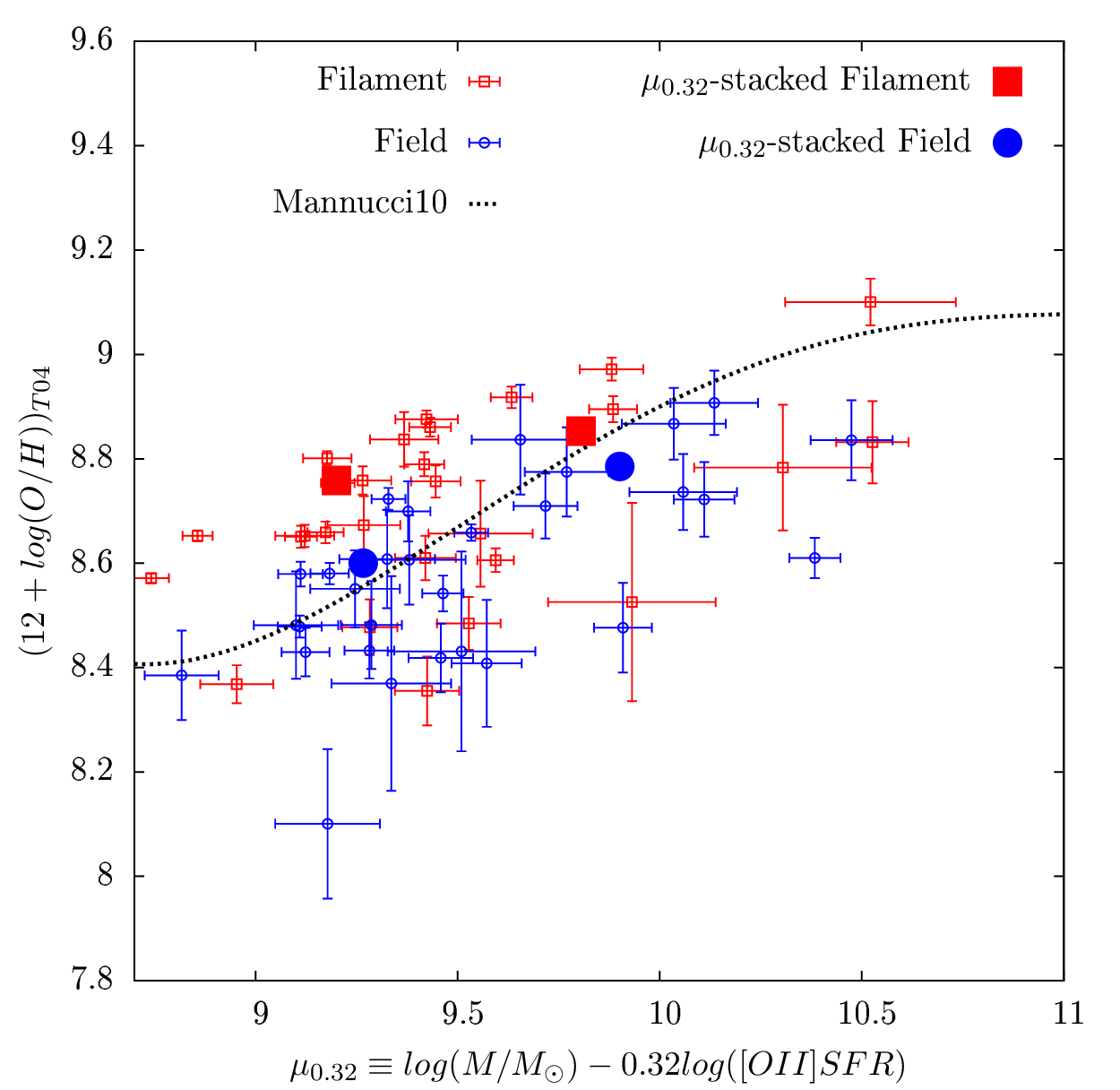}
\caption{Relation between metallicity and $\mu_{0.32}\equiv$ log(M/M$_{\odot}$)$-$0.32log([OII] SFR) for filament (red empty points) and field galaxies (blue empty points), together with $\mu_{0.32}$-stacked filament (red solid points) and field (blue solid points) spectra. For comparison, the polynomial best fit from \cite{Mannucci10} is also shown. Both filament and field galaxies nearly follow the \cite{Mannucci10} trend. However, at a fixed $\mu_{0.32}$, filament galaxies are more metal-enriched ($\sim$0.07$-$0.16 dex). The stacked data errorbars are smaller than the size of the points.}
\label{fig:MZSFR}
\end{center}
\end{figure}
   
In another attempt to quantify the environmental dependence of the mass-metallicity relation, we use the generalized two-dimensional K-S test \citep{Fasano87} to determine the probability that the populations of filament and field star-forming galaxies on the mass-metallicity plane are drawn from the same parent distribution. The derived p-value is p=0.04, implying that it is $<$ 4\% probable that the difference in mass-metallicity distributions in different environments is due to chance. In addition, we perform the one-dimensional K-S test for stellar mass and metallicity distributions in different environments. The derived p-value for the K-S test of stellar mass distributions in filament and the field is p=0.27, whereas it is p=0.03 for metallicity distributions. This indicates that it is unlikely that the stellar masses in filament and the field are drawn from a different parent distribution while there is a more significant difference between metallicity distributions.

As we already mentioned, the mass-metallicity relation is due to a more general mass-metallicity-SFR relation (the fundamental metallicity relation, \citealp{Mannucci10}). As shown in \cite{Mannucci10}, by introducing the quantity $\mu_{0.32}$=log(M)$-$0.32log(SFR) that minimizes the dispersion of the projected fundamental metallicity relation, all galaxies out to $z\sim$2.5 should follow the same $\mu_{0.32}$-metallicity relation. We define $\mu_{0.32}$=log(M/M$_{\odot}$)-0.32log([OII] SFR) and investigate the $\mu_{0.32}$-metallicity relation for filament and field galaxies. Figure \ref{fig:MZSFR} shows the relation between metallicity and $\mu_{0.32}$ for individual filament and field galaxies, along with stacked points. For comparison, the polynomial best fit from \cite{Mannucci10} is also shown. Both filament and field galaxies nearly follow the \cite{Mannucci10} trend. However, at a fixed $\mu_{0.32}$, filament galaxies are more metal-enriched. For example, at $\mu_{0.32}$ $>$ 9.5, the stacked spectra in the filament are about $\sim$0.07 dex more metal-enhanced and at $\mu_{0.32}$ $<$ 9.5, filament galaxies show a metallicity enhancement of $\sim$ 0.16 dex compared to the field galaxies. The metallicities for the stacked spectra in different environments are presented in Table \ref{stack}.

Our result regarding a deficit of metallicity in both filament ($\sim$0.06$-$0.13 dex) and field ($\sim$0.21$-$0.23 dex) galaxies at intermediate redshifts ($z\sim$0.5) compared to $z\sim$0, is in an overall qualitative agreement with \cite{Savaglio05} (at least 0.1 dex at $z\sim$0.7), \cite{Cowie08} ($\sim$0.2 dex at $z\sim$0.77 for log(M/M$_{\odot}$)=10$-$11 galaxies), \cite{Maiolino08} ($\sim$0.1$-$0.2 dex at $z\sim$0.7 for log(M/M$_{\odot}$)=9$-$10 galaxies), \cite{Lamareille09} ($\sim$0.2 dex at $z\sim$0.59), \cite{Moustakas11} ($\sim$0.16 dex at 0.45$<z<$0.55 for log(M/M$_{\odot}$)=10.1$-$10.3 galaxies), \cite{Zahid11} (up to $\sim$0.15 dex at $z\sim$0.8 for log(M/M$_{\odot}$) $<$ 10.5 galaxies), \cite{Lara-lopez13} ($\sim$0.1 dex at $z\sim$0.36), and \cite{Perez-montero13} (0.14 dex at 0.4$<z<$0.6). Qualitatively, our filament mass-metallicity relation evolution for stacked spectra resembles that of \cite{Moustakas11} and \cite{Zahid11}, whereas for field galaxies, it is similar to that of \cite{Lamareille09}.
 
The metallicity enhancement ($\sim$0.1$-$0.15 dex) in filament star-forming galaxies with respect to those in the field is consistent with several studies at low-$z$ ($z\lesssim$0.2, \citealp{Shields91,Skillman96,Mouhcine07,Cooper08,Ellison09,Scudder12,Petropoulou12,
Hughes13,Alpaslan15,Sobral15}), intermediate-$z$ (z$\sim$1, \citealp{Sobral13}), and high-$z$ (z$\sim$2, \citealp{Kulas13,Shimakawa15,Kacprzak15}), showing no or at best a small metallicity enrichment ($\sim$ 0.05$-$0.2 dex) of galaxies in clusters relative to the field or galaxies in higher local densities with respect to those in low-density environments. Our result, together with these studies suggest that the internal processes are the primary driver of the chemical evolution of star-forming galaxies and the environment has a secondary role in the metallicity evolution of galaxies. However, we stress that some studies have found opposite environmental trends in the mass-metallicity relation. We will discuss it later in Section \ref{dis5}. 

\subsection{Ionization Parameter} \label{IP}

The ionization parameter ($q$) ---a measure of the number of H-ionizing photons per H atom--- can be estimated using the O$_{32}$ index defined as the [OIII]/[OII] line ratio \citep{Kewley02,Nakajima13}. The [OIII]/[OII] ratio also depends strongly on metallicity (and stellar mass according to the mass-metallicity relation). Here, in order to estimate the ionization parameter, we use the equation given by \cite{Kobulnicky04} (see also \citealp{Nakajima14}) based on the photoionization model of \cite{Kewley02}:
\begin{multline}
log(q)=\frac{A(O_{32},12+log(O/H))}{B(O_{32},12+log(O/H))}
\\
A=32.81-1.153y^{2} \\
+(12+log(O/H))(-3.396-0.025y+0.1444y^{2})
\\
B=4.603-0.3119y-1.163y^{2} \\
+(12+log(O/H))(-0.48+0.0271y+0.02037y^{2})
\end{multline}
where $y$=log(O$_{32}$), O$_{32}$ is the [OIII]/[OII] line ratio, i.e., O$_{32}$=([OIII]$\lambda$4959+[OIII]$\lambda$5007)/[OII]$\lambda$3727, and 12+log(O/H) is the gas phase oxygen abundance (metallicity). 12+log(O/H) metallicity is estimated in Section \ref{met} using the R$_{23}$ index and the equation given in \cite{Tremonti04}. The error of ionization parameter is estimated using the Monte-Carlo approach explained in Section \ref{ED}, based on the uncertainties in estimated metallicity and the [OII] and [OIII] line fluxes. Line fluxes are estimated by fitting Gaussian functions to the line profiles. We note that the [OII] line is double-peaked and a better estimate of the flux can be achieved by fitting a double-peaked Gaussian function similar to Section \ref{ED}. However, we found the difference between the flux estimated from single and double Gaussian functions to be negligible and use the single Gaussian fit to the [OII] line in this section.

\begin{table*}
\begin{center}
\caption[Properties of the Stacked Spectra]{Properties of the Stacked Spectra for Different Subsamples}
\centering
\begin{tabular}{lcccccccc}
\hline
Subsample & [OII]3726/[OII]3729 & $N_{e}$ & [OIII]/[OII] & log($q$) & (12+log(O/H))$_{T04}$ & EW$_{[OII]}$ \\
& & cm$^{-3}$ & & log(cms$^{-1}$) & & {\AA} \\
\hline
Filament all & 0.6909$\pm$0.0044 & 22$\pm$4 & 0.3897$\pm$0.0079 & 7.352$\pm$0.086 & 8.766$\pm$0.004 & 30.4$\pm$0.7\\
log(M/M$_{\odot}$)=9$-$9.5 & 0.7273$\pm$0.0054 & 58$\pm$5 & 0.7446$\pm$0.0163 & 7.574$\pm$0.085 & 8.708$\pm$0.004 & 33.3$\pm$0.9 \\
log(M/M$_{\odot}$)=9.5$-$10 & 0.6399$\pm$0.0065 & $<$10 & 0.2457$\pm$0.0059 & 7.204$\pm$0.113 & 8.766$\pm$0.005 & 23.5$\pm$0.6 \\
log(M/M$_{\odot}$)$>$10 & 0.7008$\pm$0.0169 & 31$\pm$16 & 0.1399$\pm$0.0072 & 7.091$\pm$0.279 & 8.890$\pm$0.005 & 13.3$\pm$0.2 \\
log(SFR)$\leq$0.75 & 0.7322$\pm$0.0082 & 63$\pm$8 & --- & --- & 8.757$\pm$0.005 & --- \\
log(SFR)$>$0.75 & 0.6737$\pm$0.0042 & $<$10 & --- & --- & 8.796$\pm$0.003 & --- \\
log(sSFR)$\leq -$9 & 0.7449$\pm$0.0132 & 76$\pm$14 & --- & --- & --- & 17.0$\pm$0.4 \\
log(sSFR)$>-$9 & 0.6847$\pm$0.0043 & 15$\pm$4 & --- & --- & --- & 41.4$\pm$0.8 \\
$\mu_{0.32}\leq$9.5 & --- & --- & --- & --- & 8.759$\pm$0.003 & --- \\
$\mu_{0.32}>$9.5 & --- & --- & --- & --- & 8.853$\pm$0.004 & --- \\
\hline
Field all & 1.0050$\pm$0.0104 & 367$\pm$13 & 0.5374$\pm$0.0135 & 7.397$\pm$0.099 & 8.639$\pm$ 0.007 & 27.6$\pm$0.7 \\
log(M/M$_{\odot}$)=9$-$9.5 & 0.8621$\pm$0.0094 & 201$\pm$10 & 0.9496$\pm$0.0258 & 7.542$\pm$0.097 & 8.541$\pm$0.007 & 39.0$\pm$1.6 \\
log(M/M$_{\odot}$)=9.5$-$10 & 1.2822$\pm$0.0188 & 741$\pm$28 & 0.3413$\pm$0.0074 & 7.271$\pm$0.093 & 8.684$\pm$0.006 & 26.2$\pm$0.5 \\
log(M/M$_{\odot}$)$>$10 & 1.0105$\pm$0.0234 & 374$\pm$28 & 0.2464$\pm$0.0143 & 7.223$\pm$0.255 & 8.804$\pm$0.007 & 15.6$\pm$0.2 \\
log(SFR)$\leq$0.75 & 0.9055$\pm$0.0099 & 250$\pm$11 & --- & --- & 8.595$\pm$0.006 & --- \\
log(SFR)$>$0.75 & 1.0528$\pm$0.0139 & 427$\pm$18 & --- & --- & 8.763$\pm$0.004 & --- \\
log(sSFR)$\leq -$9 & 0.9074$\pm$0.0115 & 252$\pm$13 & --- & --- & --- & 20.7$\pm$1.0 \\
log(sSFR)$>-$9 & 1.0354$\pm$0.0117 & 405$\pm$15 & --- & --- & --- & 33.5$\pm$0.5 \\
$\mu_{0.32}\leq$9.5 & --- & --- & --- & --- & 8.600$\pm$0.004 & --- \\
$\mu_{0.32}>$9.5 & --- & --- & --- & --- & 8.785$\pm$0.006 & --- \\
\hline
\label{stack}
\end{tabular}
\end{center}
\end{table*}

\begin{figure}
\begin{center}
\includegraphics[width=3.5in]{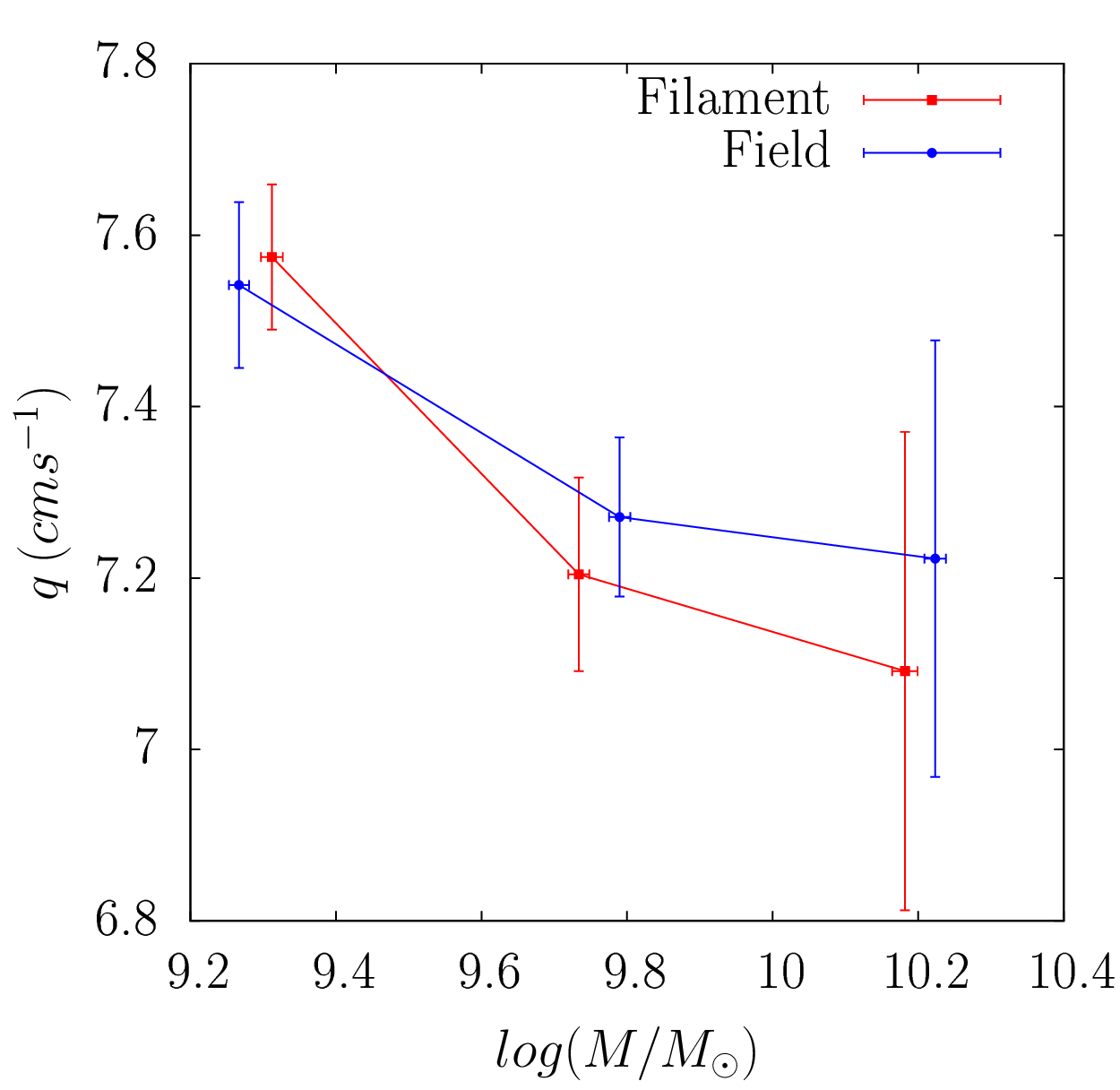}
\caption{Ionization parameter estimated with O$_{32}$ index for filament (red) and field (blue) galaxies as a function of stellar mass. Ionization parameter decreases with increasing stellar mass. However, within the uncertainties, there is no significant difference between ionization parameter of filament and field galaxies.}
\label{fig:ion}
\end{center}
\end{figure}

Within the uncertainties, we do not find a significant difference between the average ionization parameter of the star-forming galaxies in the filament compared to that of the field galaxies. The ionization parameter is estimated to be $q$=7.352$\pm$0.086 and 7.397$\pm$0.099 cms$^{-1}$ for stacked filament and field star-forming galaxies, respectively. We find that the ionization parameter is a function of stellar mass (or metallicity) and decreases with increasing stellar mass (metallicity). However, at a fixed stellar mass, we do not find a significant difference between the average ionization parameter of galaxies in the filament and the field (Figure \ref{fig:ion}). The O$_{32}$ values and the ionisation parameters for different environments and stellar mass bins are given in Table \ref{stack}. 

Recently, \cite{Sobral15} studied a merging cluster at $z\sim$0.2 and did not find a significant difference between ionisation parameter of H$\alpha$ star-forming galaxies in the field, the core, and the outskirts of the cluster at a fixed metallicity. \cite{Kewley15} also compared the ISM of a sample of field and cluster star-forming galaxies at $z\sim$2 and found no significant difference between their ionisation parameter. These are consistent with what we found in this section for our filament and field star-forming galaxies. 

\subsection{Line Equivalent Width} \label{EW}

In this section, we study the equivalent width (EW) of [OII]$\lambda$3727, as well as that of H$\beta$ in the filament and the field to investigate any environmental dependence. We also investigate the relation between EW, stellar mass, and sSFR and their dependence on environment.

We do not find a significant difference between EW distributions in different environments. The median [OII] EW for filament star-forming galaxies is EW$_{[OII]}$=29.9 {\AA}, fully consistent with that of the field galaxies (EW$_{[OII]}$=29.1 {\AA}). For the stacked spectra, we obtain EW$_{[OII]}$=30.4$\pm$0.7 {\AA} in the filament and EW$_{[OII]}$=27.6$\pm$0.7 {\AA} in the field. We obtain the same results based on other emission lines. For example, EW$_{H\beta}$ has a median value of EW$_{H\beta}$=6.0 and 8.2 {\AA} in filament and the field (corrected for Balmer stellar absorption, see Section \ref{SFR}), correspondingly, with typically larger EW uncertainties compared to those of EW$_{[OII]}$. We perform K-S test to compare the distribution of EWs in different environments. The p-value is p=0.62 ([OII]) and p=0.40 (H$\beta$), meaning that it is unlikely that the distribution of EWs in filament and the field are drawn from different parent distributions.

\begin{figure}
\begin{center}
\includegraphics[width=3.5in]{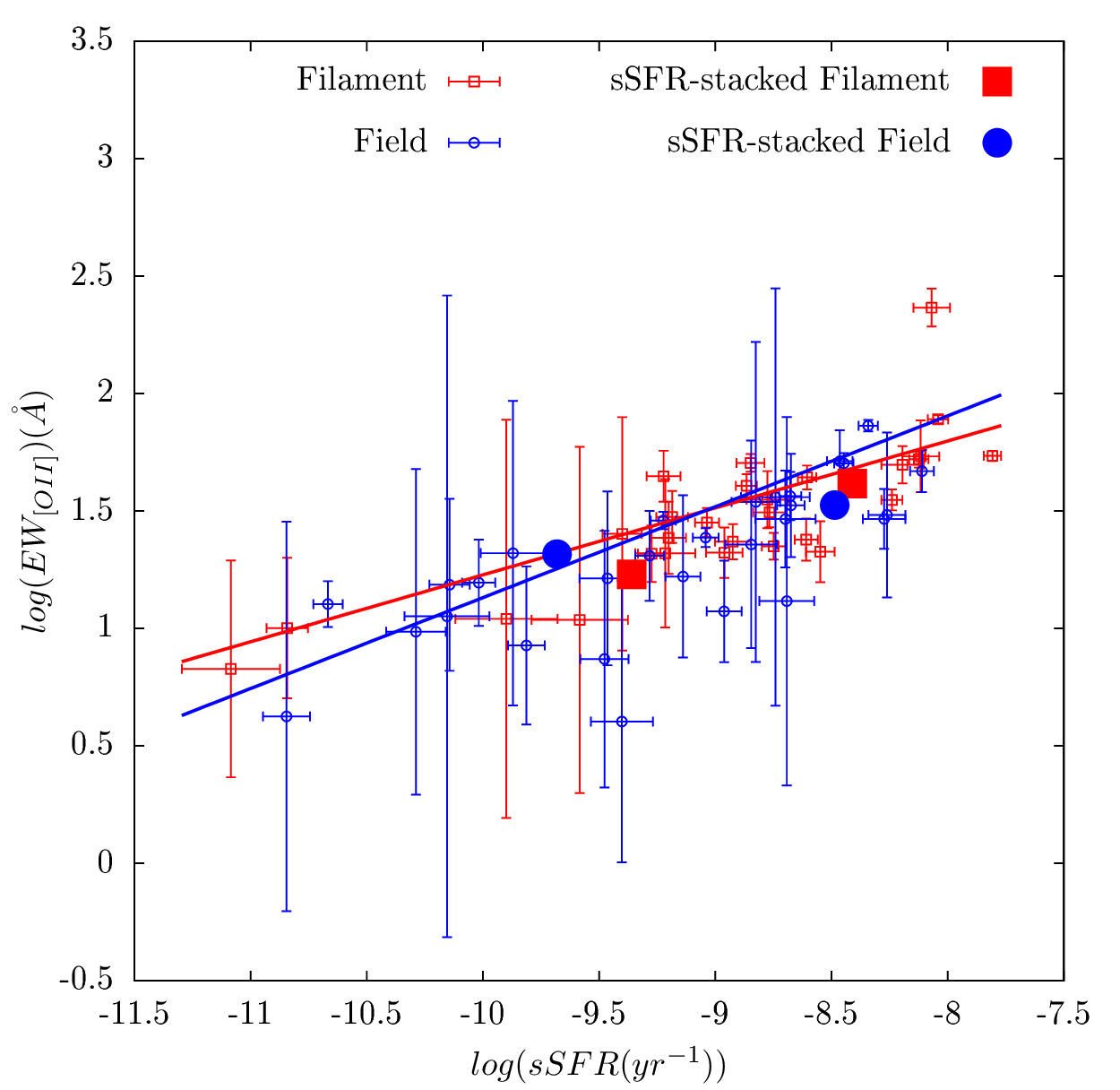}
\caption{Relation between log(EW$_{[OII]}$) and log(sSFR$_{[OII]}$) for filament (red empty points) and field (blue empty points) galaxies and the sSFR-stacked spectra (solid points). Solid lines show the linear fit to the filament (red line) and field (blue line) data points. The sSFR$_{[OII]}$ increases with EW$_{[OII]}$ for both filament and field star-forming galaxies. However, the EW$_{[OII]}$-sSFR$_{[OII]}$ relation is independent of environment, within the uncertainties. The stacked data errorbars are smaller than the size of the points.}
\label{fig:EW-sSFR}
\end{center}
\end{figure}

The EW of an emission line is a measure of its intensity with respect to the stellar continuum. Since the flux of the emitting line is related to the recent star-formation activity in galaxies and the stellar mass of a galaxy is predominantly determined by the overall mass of continuum stars, the EW of strong emission lines is a measure of the specific star-formation rate (sSFR) in galaxies \citep{Fumagalli12}. Here, we investigate the environmental dependence of the relation between EW$_{[OII]}$ and [OII] specific star-formation rate (sSFR$_{[OII]}$). Figure \ref{fig:EW-sSFR} shows log(EW$_{[OII]}$) versus log(sSFR$_{[OII]}$) for filament and field galaxies and the sSFR-stacked spectra. We find that sSFR$_{[OII]}$ increases with EW$_{[OII]}$ for both filament and field star-forming galaxies. The EW$_{[OII]}$-sSFR$_{[OII]}$ relation appears to be independent of environment. We perform a linear fit (log(EW$_{[OII]}$)=$a$[log(sSFR$_{[OII]}$)]+$b$) to the log(EW$_{[OII]}$)-log(sSFR$_{[OII]}$) relation for filament and field star-forming galaxies. Parameters of the best fit are $a$=0.285$\pm$0.064 and $b$=4.078$\pm$0.528 for filament star-forming galaxies, consistent with the field parameters ($a$=0.387$\pm$0.042 and $b$=5.001$\pm$0.364). The K-S test on the 2D distribution of galaxies in log(EW$_{[OII]}$)-log(sSFR$_{[OII]}$) plane shows similarities between filament and field galaxies (p-value=0.13).

\begin{figure}
\begin{center}
\includegraphics[width=3.5in]{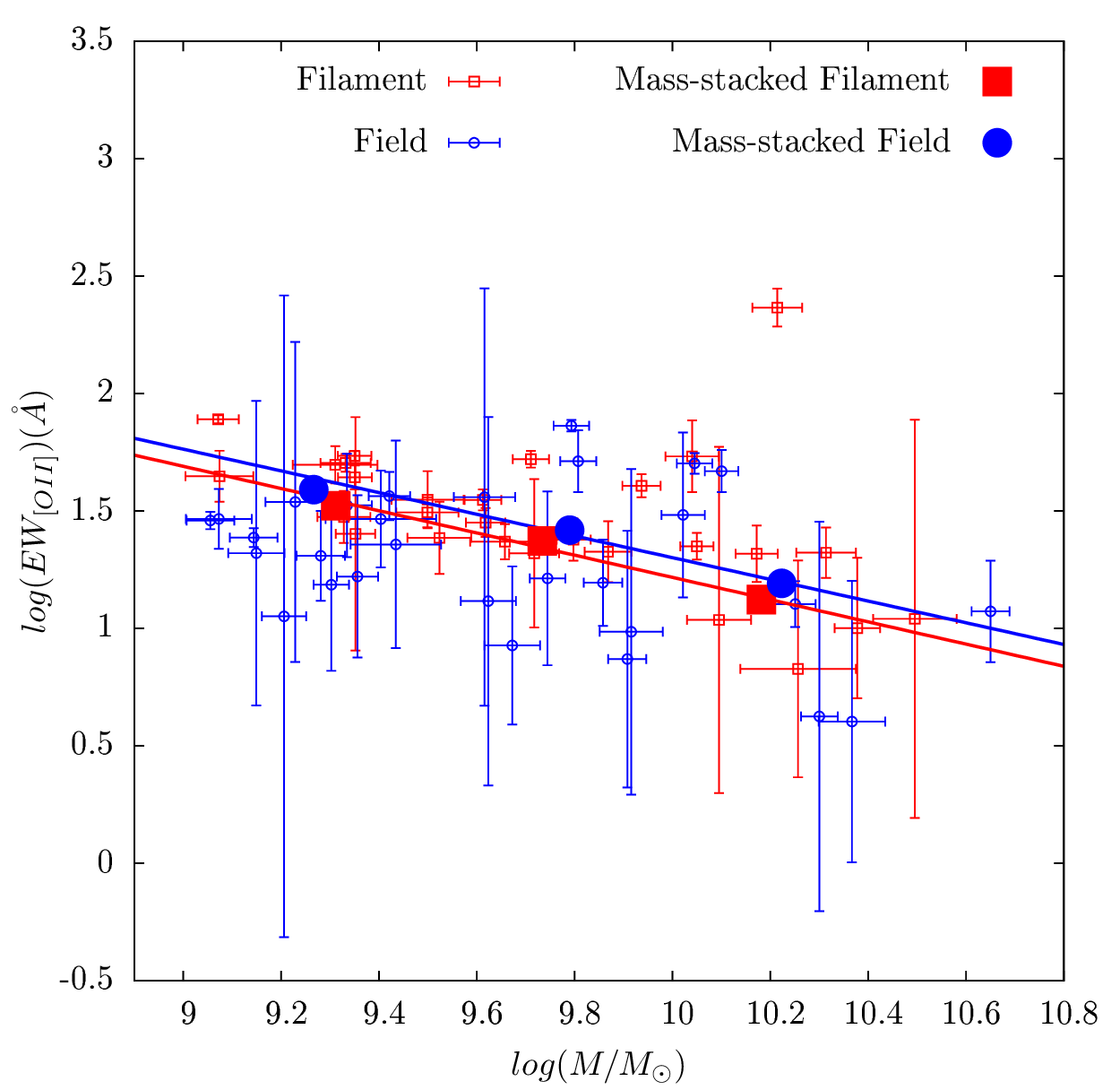}
\caption{EW$_{[OII]}$-mass anti-correlation for individual filament (red empty points) and field (blue empty points) galaxies and the stellar mass-stacked spectra (solid points). Solid lines show the linear fit to the filament (red line) and field (blue line) data. The EW$_{[OII]}$ decreases with increasing stellar mass. However, this trend is similar for star-forming galaxies in the filament and the field. The stacked data errorbars are smaller than the size of the points.}
\label{fig:EW-mass}
\end{center}
\end{figure}

Many studies have shown a relatively tight correlation (typical dispersion of $\sim$0.2 dex) between SFR and stellar mass (Main-Sequence of star-forming galaxies), or equivalently between sSFR and stellar mass out to $z\sim$6 \citep{Brinchmann04,Salim07,Daddi07,Elbaz07,Noeske07,Karim11,Whitaker12,Reddy12,Sobral14,Salmon15}.
Since we found that EW$_{[OII]}$ and sSFR$_{[OII]}$ are related, we expect to see a relation between EW$_{[OII]}$ and stellar mass. We investigate the environmental dependence of the EW$_{[OII]}$ and stellar mass for individual galaxies and our mass-stacked spectra, as shown in Figure \ref{fig:EW-mass}. We see that EW$_{[OII]}$ decreases with increasing stellar mass. However, this trend looks similar for star-forming galaxies in the filament and the field. We linearly parameterize this relation (log(EW$_{[OII]}$)=$c$[log(M)]+$d$) by fitting the stacked spectra, showing that the parameters are the same for filament ($c$=$-$0.473$\pm$0.055 and $d$=5.949$\pm$0.547) and the field galaxies ($c$=$-$0.462$\pm$0.064 and $d$=5.921$\pm$0.638), within the uncertainties.

The EW$_{[OII]}$-mass anti-correlation found here is consistent with other similar studies \citep{Fumagalli12,Sobral14,Bridge15,Cava15}. Using a sample of [OII] emitters at $z\lesssim$0.6, \cite{Bridge15} found a factor of $\sim$3 decline in the EW$_{[OII]}$ between their lowest and highest mass bin. Here, we find a factor of $\sim$2.5 decline between our lowest and highest mass bin, consistent with \cite{Bridge15} given their wider mass range at $z\sim$0.5. \cite{Fumagalli12} also found that galaxies in the lowest mass bin of their data, have on average, an EW$_{H\alpha}$ which is five times higher than those in their highest mass bin. Given their broader mass range compared to our study, it is qualitatively in agreement with our result. \cite{Sobral14} found a slope of $\sim -$0.25 between log(EW$_{H\alpha+[NII]}$) and log(M), shallower than what we found. However, this might be due to differences in selection functions and EW$_{H\alpha+[NII]}$ versus EW$_{[OII]}$. Nonetheless, our work agrees qualitatively with that of \cite{Sobral14}. We highlight that we found that the EW$_{[OII]}$-mass anti-correlation is independent of environment and it is another manifestation of the environmental invariance of the SFR$-$mass relation for star-forming galaxies, consistent with \cite{Darvish14}.

The environmental invariance of EW$_{[OII]}$-sSFR$_{[OII]}$ and EW$_{[OII]}$-mass relations found here is consistent with several observational studies and large-scale hydrodynamical simulations, showing that on average, many properties of star-forming galaxies that are related to star-formation activity (such as color, SFR, sSFR and SFR$-$mass relation) are independent of their host environment, a result that appears to hold regardless of the redshift, the method used to define environment, and the star-formation activity indicator \citep{Peng10,Wijesinghe12,Muzzin12,Koyama13,Koyama14,Hayashi14,Lin14,Vogelsberger14,Darvish14}. Hence, it has been argued that the major role of environment is to control the fraction of quiescent and star-forming systems \citep{Peng10,Sobral11,Muzzin12,Vogelsberger14,Darvish14}, that is, it is more likely for galaxies to become quenched in denser environments. However, when they form stars, on average, their star-formation activity becomes independent of their host environment. In Section \ref{dis5}, we will discuss the possible role of star-formation histories in determining the fraction of star-forming galaxies in different environments.
  
\subsection{Velocity Dispersion and Dynamical Mass} \label{VD}

In order to estimate the line-of-sight gas velocity dispersion, we use the sigma of the fitted Gaussian functions to H$\beta$ and [OIII]$\lambda$5007 line profiles. We need to subtract in quadrature the instrument velocity dispersion ($\sigma_{inst}$). The line-of-sight gas velocity dispersion is therefore:
\begin{equation}
\sigma=\sqrt{\sigma_{t}^2-\sigma_{inst}^2}
\end{equation}
where $\sigma$ is the line-of-sight gas velocity dispersion (H$\beta$ or [OIII]$\lambda$5007), $\sigma_{t}$ is the directly measured (total) velocity dispersion from the spectra (using the sigma of the Gaussian fit to the line profiles), and $\sigma_{inst}$ is the instrument velocity dispersion, measured from the width of the Arc lines. For the DEIMOS instrument (slit width 0.75", central wavelength $\lambda_{c}$=7200 {\AA}), we achieve a resolution of $\sim$1.5 {\AA}, which is equivalent to $\sigma_{inst}$=62 kms$^{-1}$. For galaxies with $\sigma_{t}$ $<$ $\sigma_{inst}$, we use the instrument velocity dispersion as an upper limit for the velocity dispersion. We note that the velocity dispersions measured from the line width have some other contributions, such as the thermal broadening due to temperature of star-forming regions (at T$\sim$10$^{4}$ K, it is $\lesssim$10 kms$^{-1}$ for H$\beta$ and much smaller for [OIII]$\lambda$5007), mergers, and turbulent gas motions inside the HII regions (typically $\sim$ 20 kms$^{-1}$, \citealp{Shields90}). We do not subtract these smaller effects from the measured velocity dispersion and the measured values are a combination of these.

\begin{figure}
\begin{center}
\includegraphics[width=3.5in]{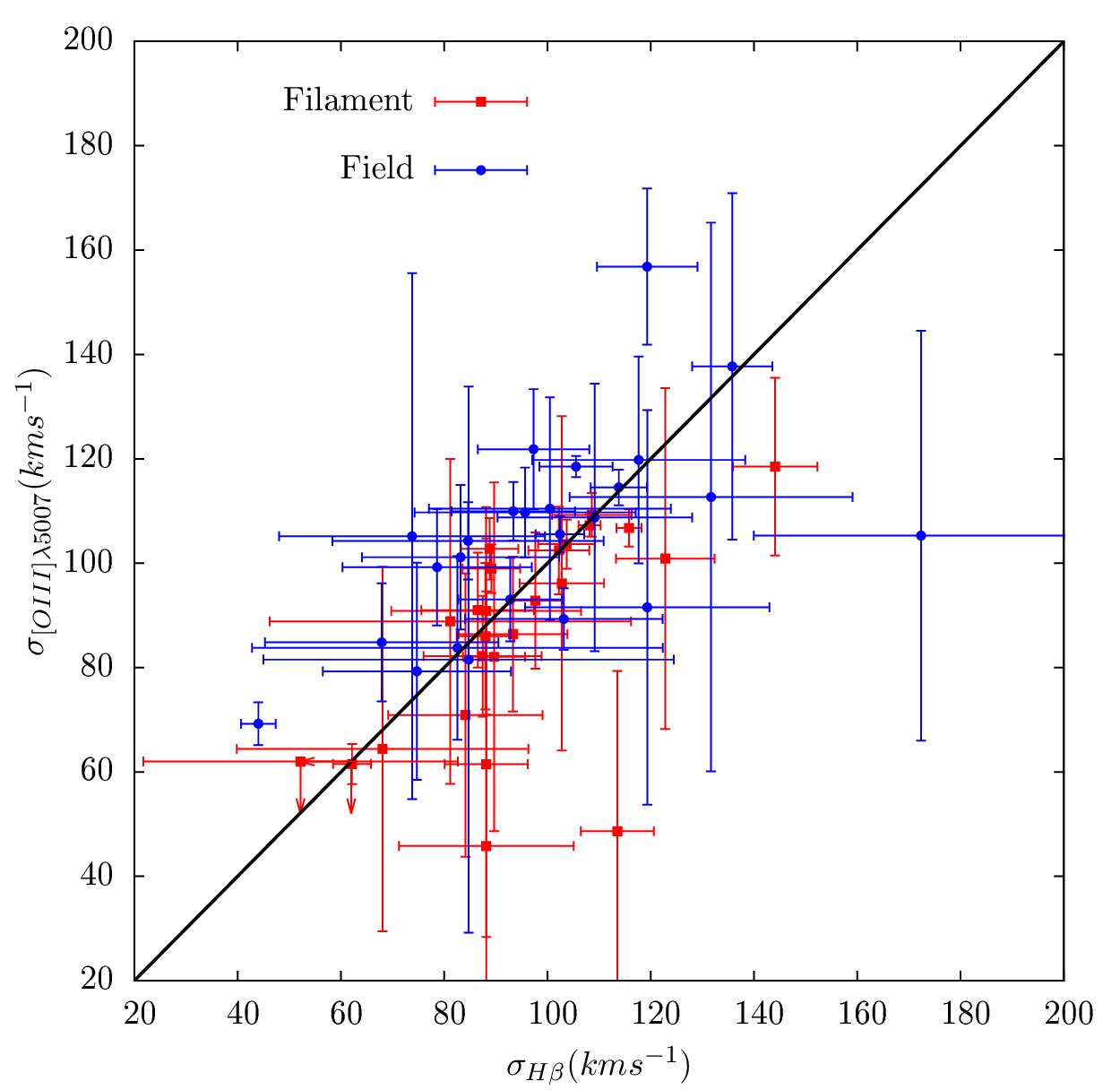}
\caption{[OIII]$\lambda$5007 versus H$\beta$ line-of-sight velocity dispersions for star-forming galaxies in the filament (red points) and the field (blue points). There is a very good agreement between the estimated velocity dispersions. The two velocity dispersion measurements have a median absolute deviation of $\sim$ 5 (15) kms$^{-1}$ in filament (field), smaller than the typical velocity dispersion uncertainties. We do not find a significant difference between velocity dispersions of the filament and field galaxies.}
\label{fig:VD}
\end{center}
\end{figure}

Figure \ref{fig:VD} shows a comparison between estimated H$\beta$ and [OIII]$\lambda$5007 velocity dispersions for star-forming galaxies in the filament and the field. There is a very good agreement between the estimated velocity dispersions. The median $\sigma_{H\beta}$ and median $\sigma_{[OIII]\lambda5007}$ are consistent to within $\sim$5 (10) kms$^{-1}$ for filament (field) galaxies, which are smaller than the median observational uncertainties of H$\beta$ velocity dispersion ($\Delta\sigma_{H\beta}\sim$10 (20) kms$^{-1}$ for filament (field) galaxies) and that of [OIII]$\lambda$5007 ($\Delta\sigma_{[OIII]\lambda5007}\sim$25 (15) kms$^{-1}$ for filament (field) galaxies). The two velocity dispersion measurements also have a median absolute deviation of $\sim$ 5 (15) kms$^{-1}$, smaller than the typical velocity dispersion uncertainties.
 
We do not find a significant difference between velocity dispersions in filament and the field. For example, the median H$\beta$ velocity dispersion is $\sigma_{H\beta}\sim$90 kms$^{-1}$ in the filament, whereas it is $\sigma_{H\beta}\sim$95 kms$^{-1}$ in the field, consistent within the uncertainties. The K-S test p-values also show that it is unlikely that the distribution of $\sigma_{H\beta}$ (p=0.47), $\sigma_{[OIII]\lambda5007}$ (p=0.43), and $\sigma_{H\beta}-\sigma_{[OIII]\lambda5007}$ 2D distribution (p=0.43) in filament and the field are drawn from different parent distributions.

We compared our estimated velocity dispersions with that of \cite{Weiner06} and found a relatively good agreement, within the uncertainties. We limited the sample of \cite{Weiner06} to typical redshift ($z\sim$0.5) and magnitude (i$^{+}\sim$21) of our own sample and found a velocity dispersion of $\sim$75 kms$^{-1}$, comparable to our estimated value of $\sim$90 kms$^{-1}$ within our typical uncertainties ($\sim$10$-$15 km$^{-1}$). Our derived $\sigma$ values also agree very well with those from \cite{Forbes96} at $\sim$0.5 ($\sim$90 kms$^{-1}$).

Given the velocity dispersion and the half-light radius of galaxies, we can estimate their dynamical mass (M$_{dyn}$) via:
\begin{equation}
M_{dyn}=\frac{\beta R_{e} \sigma^{2}}{G}
\end{equation}
where $\beta$ is a constant that depends on the mass distribution in a galaxy, $R_{e}$ is the effective half-light radius, and $G$ is the gravitational constant. The $\beta$ value varies between $\sim$2$-$6 \citep{Rix97,Pettini01,Erb03,Cappellari06,Toft12,Maseda13,Erb14,Rhoads14,Beifiori14} and is our major source of systematic uncertainty in dynamical mass measurements. Here, we use $\beta$=2 corresponding to the singular isothermal sphere mass profile \citep{Keeton01,Binney08}. However, we note that this could be as large as $\sim$6. Therefore, our dynamical masses are lower limits and could become higher by a factor $\sim$3. Half-light radii are extracted from the publicly available catalog of \cite{Leauthaud07} in the COSMOS field, derived by running $Sextractor$ \citep{Bertin96} on the $HST$/$ACS$ $F814W$ images \footnote{We also tried the public $ZEST$ catalog \citep{Scarlata07} which extracts structural parameters (including effective radii) from $HST$/$ACS$ $F814W$ images, using \cite{Simard98} $GIM2D$ $IRAF$ package. We found a relatively good agreement between the estimated effective radii from these two catalogs. Here, we only use the \cite{Leauthaud07} values.}. We use the H$\beta$ velocity dispersion for dynamical mass measurements because the typical uncertainties in H$\beta$ velocity dispersion is smaller than that of the [OIII]$\lambda$5007 line.

\begin{figure}
\begin{center}
\includegraphics[width=3.5in]{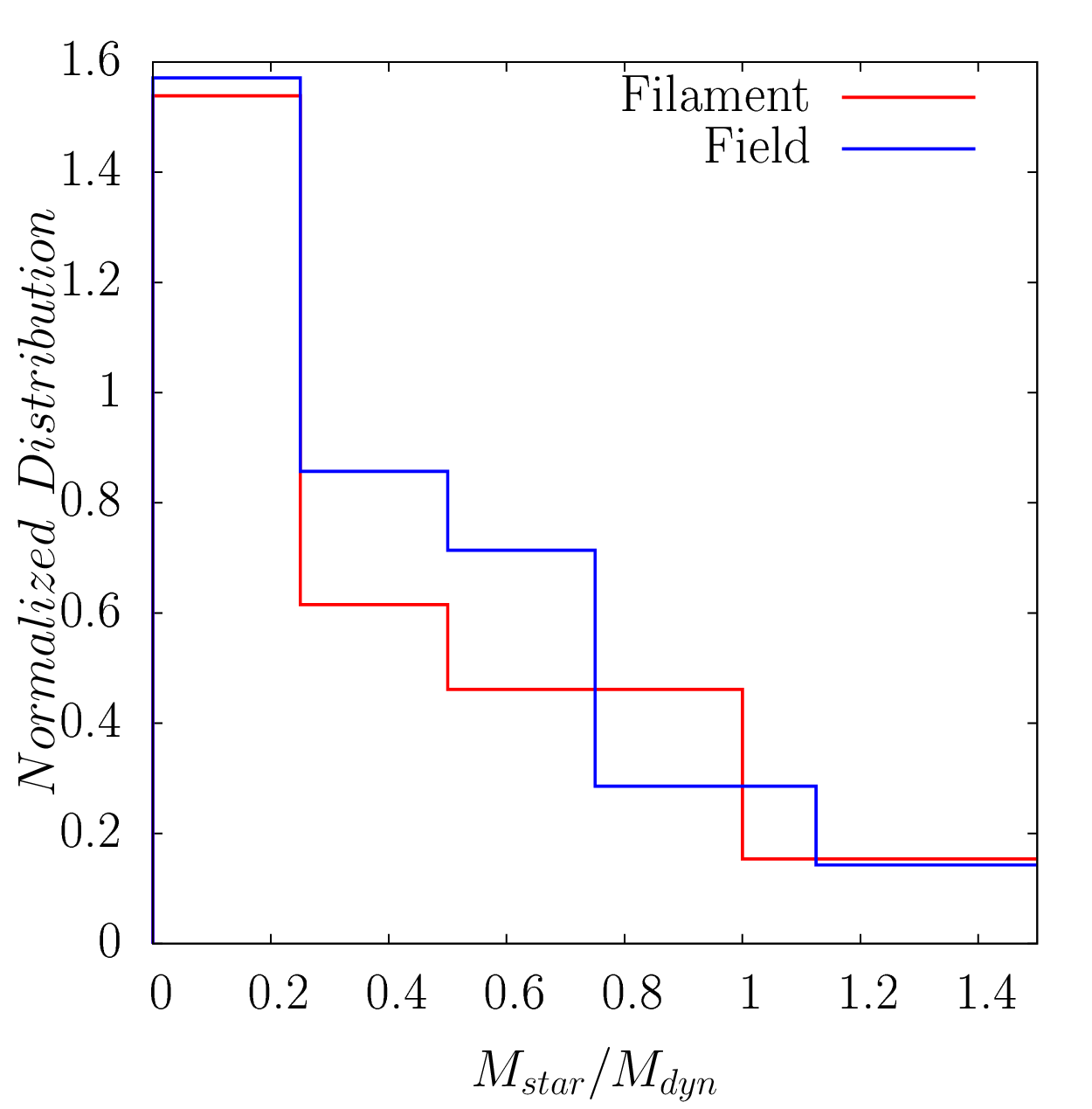}
\caption{Normalized distribution of stellar-to-dynamical mass ratio (M$_{star}$/M$_{dyn}$) for filament (red points) and the field (blue points) star-forming galaxies. We do not find a significant difference between them. After discarding the unphysical M$_{star}$/M$_{dyn}>$1 data from our analysis, we find that for filament star-forming galaxies, the median stellar-to-dynamical mass ratio is M$_{star}$/M$_{dyn}$ $\sim$0.28 with a median absolute deviation $\sim$0.15, fully consistent with the field star-forming galaxies (median M$_{star}$/M$_{dyn}$ $\sim$0.27 and median absolute deviation 0.15). The K-S test also shows no significant difference between the distribution of M$_{star}$/M$_{dyn}$ in filament and the field (p-value=0.47).}
\label{fig:DM}
\end{center}
\end{figure}

We find no significant evidence for different dynamical mass distributions for filament and field galaxies. The median dynamical mass is log(M$_{dyn}$/M$_{\odot}$)$\sim$10.12 in filament while in the field, we have median log(M$_{dyn}$/M$_{\odot}$)$\sim$10.06, with a median absolute deviation of $\sim$0.12 dex in both environments. Dynamical masses are in the range log(M$_{dyn}$/M$_{\odot}$)$\sim$9.8$-$10.4 for filament galaxies and in the range log(M$_{dyn}$/M$_{\odot}$)$\sim$9.4$-$10.8 for galaxies in the field. Dynamical masses for field galaxies have a slightly broader range. However, this might be due to a slightly wider stellar mass range in the field.

We also investigate the stellar-to-dynamical mass ratios (M$_{star}$/M$_{dyn}$) of star-forming galaxies in different environments and do not find a significant difference between them (Figure \ref{fig:DM}). For star-forming galaxies in the filament, the median stellar-to-dynamical mass ratio is M$_{star}$/M$_{dyn}$ $\sim$0.45 with a median absolute deviation $\sim$0.31. The median ratio is M$_{star}$/M$_{dyn}$ $\sim$0.30 in the field with a median standard deviation of $\sim$0.18. Given the relatively large absolute deviations, the stellar-to-dynamical mass ratios in filament and the field are consistent with each other.

Stellar-to-dynamical mass ratios in the filament are in the range M$_{star}$/M$_{dyn}$ $\sim$0.08$-$3.88, and M$_{star}$/M$_{dyn}$ $\sim$0.08$-$4.42 for field galaxies. $\sim$20\% of star-forming galaxies in the filament and $\sim$15\% in the field have unphysical M$_{star}$/M$_{dyn}>$1 (Figure \ref{fig:DM}). There might be several reasons for these unphysical values. Firstly, the uncertainties in the estimated M$_{star}$/M$_{dyn}$ are quite large and the lower limit of their estimated M$_{star}$/M$_{dyn}$ lies within the reasonable range for the majority of these sources. Secondly, we assumed a value of $\beta$=2 in estimating the dynamical masses. These sources might have a larger $\beta$ value. Choosing $\beta$ $\sim$5$-$6 brings all (except one with large error bar) the Stellar-to-dynamical mass ratios to the safe zone. Thirdly, the nebular gas dynamics may not be a complete representative of the total mass as the star-forming regions are mostly localized to inner regions of the disk. 

In fact, if we discard the seemingly unphysical M$_{star}$/M$_{dyn}>$1 from our analysis, we find a much better agreement between stellar-to-dynamical mass ratios in filament and the field. The median stellar-to-dynamical mass ratio becomes M$_{star}$/M$_{dyn}$ $\sim$0.28 (median absolute deviation $\sim$ 0.15) for filament star-forming galaxies, fully consistent with the field star-forming galaxies with median M$_{star}$/M$_{dyn}$ $\sim$0.27 and median absolute deviation 0.15. We also perform K-S test for the distributions of M$_{star}$/M$_{dyn}$ in filament and the field, showing that it is unlikely (p-value=0.47) they are drawn from different parent distributions.

The estimated stellar-to-dynamical mass ratios in the filament and the field are consistent with each other (M$_{star}$/M$_{dyn}$=0.27$-$0.28) and with some previous studies, although these studies may cover a different stellar mass range, galaxy type, redshift, and selection function when compared to our study. For example, \cite{Maseda13} found that for a sample of extreme emission line galaxies at $z\sim$2 with stellar masses log(M/M$_{\odot}$)$\sim$8$-$9, the stellar-to-dynamical mass ratio is M$_{star}$/M$_{dyn}$ $\sim$0.27 (assuming $\beta$=3), in good agreement with our result. Using a sample of massive quiescent galaxies at $z>$1 and in the SDSS, \cite{Belli14} showed that the average log(M$_{star}$/M$_{dyn}$) is $-$0.13 for $z>$1 and $-$0.12 for the SDSS samples, assuming $\beta$=5. Adopting our value of $\beta$=2 in \cite{Belli14} leads to the average value of M$_{star}$/M$_{dyn}$ $\sim$0.30 for SDSS and $z>$1 samples, fully consistent with the median value of (0.27$-$0.28) for our study at $z\sim$0.5.

According to the numerical simulations of the galaxy harassment scenario, galaxies in denser environments lose their mass due to frequent interactions with other galaxies and the tidal gravitational field of the dense environment. However, the dark matter component of a galaxy experiences a more effective mass loss compared to the stellar component, as the stellar component is more bound to the gravitational potential of the galaxy (e.g., \citealp{Moore98}). Therefore, one expects a larger stellar-to-dynamical mass ratio in denser environment. Observational evidence for the environmental dependence of M$_{star}$/M$_{dyn}$ for dwarf galaxies in the cluster environment exists (e.g., \citealp{Rys14}). However, the absence of any environmental dependence of M$_{star}$/M$_{dyn}$ in our work suggests that galaxy harassment is not significantly effective on the type of galaxies (log(M$_{star}$/M$_{\odot}$)$\gtrsim$9) and in the environments (filaments and intermediate-density regions) considered in this study.
 
Many studies of the Tully-Fisher relation out to $z\sim$1 have shown no or at best very weak (for S0 galaxies and fainter systems) environmental dependence of its characteristics \citep{Sakai00,Nakamura06,Pizagno07,Jaffe11,Mocz12,Bosch13,Rawle13,Sobral13}. Since maximum rotational velocity is related to velocity dispersion and absolute magnitude is proportional to stellar mass, the environmental invariance of stellar mass and velocity dispersion seen in our work is indirectly related to the environmental independence of the Tully-Fisher relation seen in these previous studies and obliquely agrees with them.  
      
\section{Discussion} \label{dis5}

In Section \ref{result5}, we showed that some of the observable properties of filament and field star-forming galaxies such as their EW, EW-sSFR relation, velocity dispersion, dynamical mass, and ionization parameter are the same within the uncertainties. However, we found an enhancement of metallicity and a significant reduction of electron density for filament galaxies. What might be the cause of these differences?

We first need to discuss that the environmental dependence of the metallicity of star-forming galaxies may depend on the scale at which the environment is defined. At small scales (a few tens of kpc), there is substantial evidence from observations and simulations for a decrement of metallicity and an enhancement of star-formation activity in galaxy close pairs, merging, and interacting systems compared to isolated, field galaxies, mostly attributed to the interaction-induced inflow of metal-poor gas from the periphery of interacting galaxies to the center, diluting their metal content, and increasing their gas fuel for star-formation \citep{Mihos96,Kewley06,Michel08,Ellison08,Sol10,Rupke10,Perez11,Scudder12a,Ellison13,Ly14}. At intermediate group scales where galaxy interactions are more common compared to cluster and field environments \citep{Perez09,Tonnesen12} (due to a combination of (1) a lower velocity dispersion of group galaxies relative to their cluster counterparts and (2) a higher number density of group galaxies compared to the field systems, which provide an ideal condition for interactions), there is also evidence for a deficit of metals in group galaxies compared to control samples in the field \citep{Lara-lopez13a}, possibly due to a higher fraction of interacting galaxies. At larger filamentary and cluster scales, the slight metal enhancement of galaxies relative to the field might be due to (1) the inflow of already enriched interafilamentary or interacluster gas into galaxies as observations and simulations have shown a more metal enriched IGM in cluster and filament environments compared to the field \citep{Arnaud94,Aracil06,Stocke06,Stocke07,Sato07,Dave11,Cen11,Oppenheimer12}, (2) the environmental strangulation \citep{Larson80,Peng15} of low-metallicity diluting gas falling from the surrounding LSS cosmic web into galaxies, (3) the environmental ram pressure stripping \citep{Gunn72,Abadi99} of the metal-poor diluting gas in the periphery of galaxies, and (4) trapping and recycling of metal-enriched outflows due to the hotter environment of filaments and clusters \citep{Cen06,Aracil06,Werner08} compared to the field.

However, among all the possibilities considered above, the inflow of the metal-enriched intrafilamentary gas into filament galaxies can better describe our observations in this work. Strangulation and ram pressure stripping would affect the gas content of filament galaxies which would eventually result in a reduced gas reservoir for galaxies, leading to a reduction of SFR in filament galaxies. However, we showed in Section \ref{EW} that the EW$_{[OII]}$-mass relation which is related to star-formation activity in galaxies is invariant to filament and field star-forming galaxies. Moreover, ram pressure stripping is practically effective for dwarf galaxies in denser environments of massive clusters with large velocity dispersions. Given the mass range of our sample galaxies and the environment of our filament, it is unlikely that ram pressure stripping would significantly affect the gas content of our filament galaxies.  
 
The small environmental dependence of galaxy metallicity seen in this work and similar studies might be due to the selection of mostly brighter, central galaxies as highlighted by \cite{Peng14}. Using a sample of galaxies in the SDSS, \cite{Peng14} showed that once central and satellites are carefully separated, there is a strong relation between overdensity and metallicity for satellite star-forming galaxies, whereas for central system, no such correlation was found. They attributed this correlation to the inflow of more metal-enriched gas in denser environments to satellite galaxies. Similarly, \cite{Pasquali12} found that satellites have higher metallicities than equally massive centrals in the SDSS dataset.

We showed that on average, the ionization parameter is almost the same for filament and field galaxies, whereas the electron density is much lower for filament galaxies. The ionization parameter is defined as:
\begin{equation} \label{eq:EQ1-5}
q=\frac{Q(H^{0})}{4\pi r^{2}N_{H}}
\end{equation} 
\\where $Q$($H^{0}$) is the number of Hydrogen photoionizing photons per second, passing through the surface area of 4$\pi r^{2}$ with the Str\"{o}mgren radius $r$, and $N_{H}$ is the number density of Hydrogen atoms. Assuming ionization equilibrium, the photoioization rate is balanced out by the recombination rate:
\begin{equation} \label{eq:EQ2-5}
Q(H^{0})\propto r^{3} N_{H}^{2} \epsilon
\end{equation}  
where $\epsilon$ is the filling factor. Inserting equation \ref{eq:EQ2-5} into \ref{eq:EQ1-5} gives:
\begin{equation} \label{eq:EQ3-5}
q\propto r N_{H} \epsilon
\end{equation} 
If the number density of hydrogen atoms is related to electron density, given the lower electron density in filament galaxies and similar ionization parameter between field and filament systems, the Str\"{o}mgren radius and/or the filling factor of filament galaxies should change in such a way that leave the ionization parameter intact, according to equation \ref{eq:EQ3-5}. In other words, on average, filament galaxies might have bigger HII star-forming regions assuming $\epsilon$ is fixed, or the filling factor is larger for filament galaxies compared to the field assuming that $r$ is the same in these two environments, or a combination of these.
 
Different star-formation histories between filament and field star-forming galaxies can explain some of the differences seen in this study. A longer, more constant star-formation timescale for filament star-forming galaxies compared to a shorter, burstier star-formation history for field star-forming galaxies is capable of sustaining a bigger HII region for a longer time. A bigger HII region reduces the pressure inside the region which leads to a lower electron density. A longer star-formation timescale for filament star-forming galaxies also suggests that at the time of observation, it is more likely to find a higher fraction of star-forming galaxies in filaments compared to the field. This is consistent with e.g., \cite{Darvish14} who found a higher fraction of H$\alpha$ star-forming galaxies in a filamentary structure at $z\sim$0.84. It is still not clear what environmental effects (if any) might cause the difference in electron density, Str\"{o}mgren radius, filling factor, and star-formation histories between filament and field galaxies. Whatever they are, they alter the ISM of galaxies in such a manner that affect the electron densities and leave the ionization parameter almost unchanged. Sub-kpc-resolution simulations in different environments are required to unveil the nature of this trend. Furthermore, with the publicly available $HST$/$ACS$ $F814W$ data in the COSMOS field, a detailed high-quality morphological analysis of filament and field galaxies can be performed. This can potentially shed more light on the origin of the environmental effects seen in this study, a possibility that can be done in another work in the future.

Since electron density measurements originate from HII regions with sizes up to a few hundred pc, a full understanding of them in different environments at intermediate and high redshifts requires sub-kpc resolution. Current space-based telescopes such as HST and ground-based integral field units (IFU) equipped with adaptive optics (AO) are able to provide information on kpc-scale star-forming regions in intermediate and high redshift galaxies (see e.g., \citealp{Swinbank12,Wuyts13,Hemmati14}). However, with the next generation of telescopes such as TMT and E-ELT, sub-kpc resolution for intermediate- and high-$z$ galaxies can be achieved through AO systems, enabling us to have a grasp of the properties of HII-regions in these galaxies located in different environments. 

\section{Summary and Conclusion} \label{Sum} 

In this paper, we spectroscopically confirm the presence of a large filamentary structure ($\sim$ 8 Mpc) in the COSMOS field at $z\sim$0.53. We extensively study the physical properties of the star-forming emission-line galaxies in this filament and compare them with a control sample of field galaxies at the same redshift. Our main results are:
\begin{enumerate}
\item We estimate the average electron density for filament and field star-forming galaxies using [OII]$\lambda$3726/$\lambda$3729 line ratio. {We find no clear relation between [OII]$\lambda$3726/$\lambda$3729 ratio (electron density) and stellar mass, SFR, and sSFR.} However, we show that the average electron density is significantly lower (a factor of $\sim$ 17) for filament star-forming galaxies compared to that of the field systems. We obtain the same result if we perform the comparison in fixed stellar mass, SFR, and sSFR bins. Depending on the bin selected, electron densities vary in the range $\sim$ $<$10$-$75 cm$^{-3}$ for filament star-forming galaxies and in the range $\sim$ 200$-$740 cm$^{-3}$ for star-forming galaxies in the field. However, we note that the dispersion in the estimated line ratios (electron densities) for individual galaxies is still large and a larger sample is required to more robustly constrain the environmental dependence of electron densities for star-forming galaxies. The decrement of the electron densities in filament star-forming galaxies might be due to a longer star-formation timescale for filament star-forming systems. 
\item We stack the galaxies based on their local overdensity value (regardless of whether they are in the filament or the field) and find an anti-correlation between the average electron density of the stacked-spectra and the local overdensity of galaxies. This suggests that the environmental dependence of the electron density for star-forming galaxies is possibly a local effect.
\item We estimate the oxygen abundance metallicity using the R$_{23}$ index. We show that our mass-metallicity relation at $z\sim$0.5 for filament and field star-forming galaxies lies in between mass-metallicity relations at $z\sim$0 and $z\sim$2. We also show that our $\mu_{0.32}$-metallicity relation for filament and field star-forming galaxies nearly follows the \cite{Mannucci10} trend at $z\sim$0.
\item We show that at a fixed stellar mass, SFR, and $\mu_{0.32}$, filament star-forming galaxies are more metal-enriched ($\sim$0.1$-$0.15 dex) compared to their field counterparts. The metallicity enhancement for filament star-forming galaxies compared to the field might be due to the inflow of the already enriched intrafilamentary gas into filament galaxies. The metallicity enhancement in denser environments suggests that environment has a secondary effect in the metal enrichment of galaxies and internal processes are the major driver of the chemical evolution in galaxies.
\item We estimate the average ionisation parameter of filament and field star-forming galaxies using the O$_{32}$ index. Ionisation parameter depends on stellar mass and decreases with increasing stellar mass for both filament and field star-forming galaxies. However, within the uncertainties, we find insignificant difference between ionisation parameter of filament and field star-forming systems.\\
\item We show that EW$_{[OII]}$ correlates with sSFR and anti-correlates with stellar mass for filament and field star-forming galaxies. However, we show that within the uncertainties, EW$_{[OII]}$-sSFR and EW$_{[OII]}$-mass relations are the same for filament and field star-forming systems. The environmental independence of the EW$_{[OII]}$-mass anti-correlation is another manifestation of the environmental invariance of the SFR$-$mass relation for star-forming galaxies, seen in previous studies. 
\item We estimate the line-of-sight velocity dispersion and dynamical masses for filament and field star-forming systems. We show that, on average, velocity dispersion, dynamical mass, and stellar-to-dynamical mass ratio are similar for filament and field star-forming galaxies. The environmental invariance of stellar-to-dynamical mass ratio in this work suggests that galaxy harassment is not significantly effective on the type of galaxies (log(M$_{star}$/M$_{\odot}$)$\gtrsim$9) and in the environments (filaments and intermediate-density regions) considered in this study.
\end{enumerate}
Assuming ionisation equilibrium within HII star-forming regions, we can show that the ionisation parameter directly depends on the electron density, the diameter of the HII region, and the volume filling factor. Since we showed that, on average, the ionisation parameter is similar for filament and field star-forming galaxies and the electron density is much lower for filament star-forming systems, this suggests that, on average, filament star-forming galaxies might have larger HII regions compared to their field counterparts if we assume that the filling factor is the same for them. However, HII regions have sub-kpc sizes and a full understanding of their sizes in different environments at higher redshifts requires the next generation of telescopes such as TMT that are equipped with the IFU systems and assisted by the AO technology.     

\section*{acknowledgements}
We gratefully thank the anonymous referee for reading the manuscript and providing very useful comments which significantly improved the quality of this work. The authors wish to thank Brian Siana, Gabriela Canalizo, Brant Robertson and Michael Rich for their helpful comments and discussions. DS acknowledges financial support from the Netherlands Organisation for Scientific research (NWO) through a Veni fellowship, from FCT through a FCT Investigator Starting Grant and Startup Grant (IF/01154/2012/CP0189/CT0010), from FCT grant (UID/FIS/04434/2013), and from LSF and LKBF. The observations presented herein were obtained at the W. M. Keck Observatory, which is operated as a scientific partnership among the California Institute of Technology, the University of California and the National Aeronautics and Space Administration. The Observatory was made possible by the generous financial support of the W. M. Keck Foundation. The authors would like to recognize and acknowledge the very prominent cultural role and reverence that the summit of Mauna Kea has always had within the indigenous Hawaiian community. We are fortunate to have the opportunity to perform observations from this mountain.
\bibliographystyle{apj} 
\bibliography{fil-6-new}

\end{document}